\documentclass[%
 reprint,
 amsmath,amssymb,
 aps,
 prd,
 floatfix,
 dvipsnames,
]{revtex4-2}

\usepackage{graphicx} 
\usepackage{dcolumn} 
\usepackage{hyperref} 
\usepackage{verbatim} 
\usepackage{enumitem} 

\newcommand*{\nolink}[1]{%
  \begin{NoHyper}#1\end{NoHyper}%
}

\usepackage[caption=false]{subfig}

\usepackage{bm} 
\usepackage{bbm} 
\usepackage{braket} 
\usepackage{tensor} 
\usepackage{siunitx} 
\usepackage{tikz-cd} 

\usepackage{acro}

\DeclareAcronym{udw}{
    short = UDW,
    long = Unruh-DeWitt
}
\DeclareAcronym{qft}{
    short = QFT,
    long = quantum field theory
}
\DeclareAcronym{com}{
    short = c.m.,
    long = center of mass
}
\DeclareAcronym{qed}{
    short = QED,
    long = quantum electrodynamics
}
\DeclareAcronym{rqm}{
    short = RQM,
    long = relativistic quantum mechanics
}
\DeclareAcronym{nrqm}{
    short = NRQM,
    long = non-relativistic quantum mechanics
}
\DeclareAcronym{mus}{
    short = MUS,
    long = minimum uncertainty states
}
\DeclareAcronym{rmus}{
    short = RMUS,
    long = relativistic minimum uncertainty states
}
\DeclareAcronym{fw}{
    short = FW,
    long = Foldy-Wouthuysen
}
\DeclareAcronym{shp}{
    short = SHP,
    long = Stueckelberg-Horwitz-Piron
}

\begin{document}

\preprint{APS/123-QED}

\title{Relativistic Unruh-DeWitt detectors with quantized center of mass}

\author{Evan P. G. Gale}
\email{e.gale@uq.edu.au}

\author{Magdalena Zych}%
\email{m.zych@uq.edu.au}

\affiliation{%
    ARC Centre of Excellence for Engineered Quantum Systems, School of Mathematics and Physics, The University of Queensland, St Lucia, QLD 4072, Australia
}%

\date{\today}

\begin{abstract}
In this paper, we extend the~\acf{udw} model to include a relativistic quantized~\acf{com} for the detector, which traditionally has a classical~\ac{com} and follows a classical trajectory. We develop a relativistic model of an inertial detector following two different approaches, starting from either a first- or second-quantized treatment, which enables us to compare the fundamental differences between the two schemes. In particular, we find that the notion of localization is different between the two models, and leads to distinct predictions, which we study by comparing the spontaneous emission rates for the~\ac{udw} detector interacting with a massless scalar field. Furthermore, we consider the~\ac{udw} system in both a vacuum and medium, and compare our results to existing models describing a classical or quantized~\ac{com} at low energies. We find that the predictions of each model, including the two relativistic cases, can in principle be empirically distinguished, and our results can be further extended to find optimal detector states and processes to perform such experiments. This would clarify both the role of a quantized~\ac{com} for interactions with an external field, and the differing localizations between the first- and second-quantized treatments.

\end{abstract}

\maketitle



\section{Introduction}\label{sec:intro}

The~\acf{udw} model, originally proposed by~\citet{unruh_notes_1976} and later simplified to its monopole form by~\citet{dewitt_quantum_1979}, has become the customary model for particle detectors in quantum physics. It was originally formulated for the investigation of~\ac{qft} in both flat and curved spacetimes~\cite{takagi_vacuum_1986, birrell_quantum_1982, crispino_unruh_2008}, is commonplace in the study of relativistic quantum information~\cite{reznik_entanglement_2003,*reznik_violating_2005, olson_entanglement_2011,*olson_extraction_2012, alsing_observer-dependent_2012} and entanglement harvesting~\cite{Note1, salton_acceleration-assisted_2015, martin-martinez_sustainable_2013,*pozas-kerstjens_harvesting_2015,*pozas-kerstjens_entanglement_2016}, and is used in quantum optics as an idealized model of the atom-light interaction~\cite{del_rey_simulating_2012, martin-martinez_wavepacket_2013, lopp_quantum_2021}.

Conventionally, the detector is modeled as a quantized two-level system following a classical trajectory, i.e.,~with a classical~\acf{com}. The traditional~\ac{udw} model is an idealization which does not account for fully quantized dynamics, but can be extended to include quantized external degrees of freedom. Recently, such a detector model featuring a quantized~\ac{com} was studied in the non-relativistic regime by \nolink{\citeauthor{stritzelberger_coherent_2020}}~\cite{stritzelberger_coherent_2020,*stritzelberger_coherent_2020-1,*sudhir_unruh_2021,*stritzelberger_entanglement_2021} for inertial and uniformly accelerating detectors (but see also~\cite{guo_spontaneous_2008} for an earlier, atom-light treatment). This model was proposed in order to study the dynamics of a quantized~\ac{com}, and in particular the effects of the~\ac{com} coherence on the detector-field interaction.

However, a non-relativistic treatment of a quantized~\ac{com} does not come without issues, and raises the question of the consistency of the resulting model. These subtleties are especially relevant in the case of~\ac{qed} and quantum optics, where a quantized~\ac{com}~extension of the detector model is often considered (see, e.g.,~\cite{lopp_quantum_2021} and references within). In an early paper studying the effects of a~\ac{com} in~\ac{qed},~\citet{wilkens_spurious_1993} found spurious velocity-dependent effects due to the omission of requisite relativistic corrections, namely the R\"{o}ntgen interaction term~\cite{rontgen_ueber_1888}, which were resolved once this correction was introduced~\cite{wilkens_significance_1994}. However, the inclusion of the R\"{o}ntgen term alone is not enough to deal with all spurious effects; the mass defect of the detector must also be considered in order to avoid anomalous friction forces~\cite{sonnleitner_will_2017,*sonnleitner_mass-energy_2018}. These non-physical results are ultimately due to the mixing of the Galilean and Lorentz groups, which respectively govern the dynamics of the detector and field. The problems of the model can be partially remedied by introducing the R\"{o}ntgen term and modeling the mass defect, but as Wilkens noted, ``the ultimate cure of this deficiency may be expected from a relativistically covariant description of the atomic motion~\textellipsis~this program is, however, highly nontrivial''~\cite{wilkens_spurious_1993}.

There are two possible approaches to developing a relativistic formulation of the~\ac{com}~dynamics, one starting from a second-quantized description and the other by developing a relativistic first-quantized description. While the detector in the~\ac{udw} model is most commonly formalized as a first-quantized system, a second-quantized model was also proposed in Unruh's original paper~\cite{unruh_notes_1976}, and has since been considered in a number of recent analyses: in the study of backreaction effects on an accelerating, finite-mass detector~\cite{parentani_recoils_1995,*parentani_schwinger_1997,*gabriel_interacting_1998, casadio_accelerated_1995,*casadio_accelerated_1999} (see also~\cite{reznik_unruh_1998} for a first-quantized treatment); in the study of detector models comparing the ``bare'' and ``dressed'' states of the detector~\cite{costa_modeling_2009}; and most recently in the study of quantum reference frames~\cite{giacomini_second-quantized_2022}.



In this paper, we extend the~\ac{udw} model to incorporate a relativistic quantized~\ac{com} for an inertial detector, and compare the first- and second-quantized approaches. We find that these two models lead to distinct notions of localization, whose consequences can be studied for a given physical process, which we consider specifically in the case of spontaneous emission. We compare the first- and second-quantized models by studying their predictions for the rate of spontaneous emission, for which we obtain analytical expressions as a functional of the detector's initial state. An advantage of working within the~\ac{udw} model is that spin is neglected, which is particularly convenient for the comparison of the different localizations. In addition to questions of localizability, we also consider a detector interacting with a field in both a vacuum and medium, from which we obtain previously proposed models for the non-relativistic~\cite{stritzelberger_coherent_2020} and ``semi-relativistic''~\cite{wood_quantized_2022} regimes as limiting cases.

Often, one develops~\ac{rqm} within the framework of a relativistic~\ac{qft}, i.e.,~second-quantized approach; the status of a relativistic non-field-theoretic quantum mechanics is controversial, and is often objected to on both formal and ontological grounds, e.g.,~regarding the non-localizability and observer-dependence of particles (for some discussion on this and related topics, see~\cite{kalnay_localization_1971, davies_particles_1984, malament_defense_1996, fleming_strange_1999, halvorson_no_2002, hobson_there_2013}). While a second-quantized approach is required for a fully relativistically covariant treatment of standard quantum mechanics, our intention here is to present and explore consequences of both of these approaches. Furthermore, analogous to how~\ac{nrqm} exists as a limiting case of relativistic~\ac{qft}, a first-quantized~\ac{rqm} should also be ``embedded'' within such a theory, provided that the system remains in a low-energy regime with respect to the Compton scale, as for example evidenced by experiments such as those studying the electron's magnetic moment~\cite{hanneke_new_2008, fan_measurement_2023}.

Throughout this paper, we refer to the different formal descriptions of quantum mechanics as either ``first-'' or ``second-quantized.'' Although this terminology was originally coined based on a historical misunderstanding, we adopt this terminology to avoid potential confusion when referring to the detector and field constituents. That is, we refrain from referring to the fully covariant approach as a ``field'' theory, and instead refer to ``first-'' or ``second-quantized'' approaches, either of which may be treated relativistically.

The paper is organized as follows. In Sec.~\ref{sec:past_models}, we review the conventional~\ac{udw} model, its history and various extensions, paying close attention to how the different models are related. We review the standard point-like and smeared detector models, the quantized~\ac{com} extension, and second-quantized treatments of the detector. In Sec.~\ref{sec:rel_first_q_model}, we present our relativistic first-quantized model of the~\ac{udw} detector and compare to the second-quantized formulation, primarily with regard to the differences in localization. Additionally, we consider the detector-field system in both a vacuum and medium, and compare the relativistic first- and second-quantized models to earlier results featuring a classical, or semi-/non-relativistic~\ac{com}. In Sec.~\ref{sec:results}, we derive the spontaneous emission rate of the detector as a functional of the initial~\ac{com} wavefunction, which we take to have a Gaussian profile, and again compare our relativistic treatment to past models. Finally, we discuss our results in Sec.~\ref{sec:discussion}, where we systematically address the issues of localizability, time, and mass-energy in~\ac{rqm}, before concluding in Sec.~\ref{sec:conc}.

We consider $(3 + 1)$-dimensional Minkowski spacetime with metric signature $(+, -, -, -)$ and abbreviate $(\bm{x}, t)$ by $x$. We work in natural units $\hbar = c = 1$, although we restore units when needed for clarity.


\section{Unruh-DeWitt detector models}\label{sec:past_models}

Unruh's~\cite{unruh_notes_1976} seminal paper originally proposed two detector models: a ``particle in a box'' detector model and a second-quantized detector model. The latter is generally not studied, while the former was simplified to the well-known point-like ``monopole'' description by~\citet{dewitt_quantum_1979}, now known simply as the \ac{udw} model. While the \ac{udw} detector has become the standard model, and the one most often treated in the literature, various iterations have since been made. In this section, we review the monopole detector model and some common extensions made by past authors, and conclude by introducing and evaluating the second-quantized model, which has been the subject of varied interests~\cite{parentani_recoils_1995, casadio_accelerated_1995, costa_modeling_2009, giacomini_second-quantized_2022}.

In its most general form, the \ac{udw} model features a detector coupled to a field, with total Hamiltonian
\begin{align} \label{eq:total_hamiltonian}
    \hat{H} = \hat{H}_D + \hat{H}_F + \hat{H}_I \, .
\end{align}
For the moment, we leave the Hamiltonians for the detector~$\hat{H}_D$ and interaction~$\hat{H}_I$ unspecified, but subsequently treat them in detail. In the \ac{udw} model, a real-valued scalar field is most commonly considered, with a free Hamiltonian given by
\begin{align}
    \hat{H}_F = \int d^3k \, \omega(\bm{k}) \, \hat{a}^\dagger(\bm{k}) \, \hat{a}(\bm{k}) \, ,
\end{align}
where $\omega(\bm{k})$ denotes the dispersion relation for the field, and $\hat{a}^\dagger(\bm{k})$ and $\hat{a}(\bm{k})$ are respectively the creation and annihilation operators, which satisfy the commutation relations $[\hat{a}(\bm{k}), \hat{a}^\dagger(\bm{k}')] = \delta^{(3)}(\bm{k} - \bm{k}')$.

The scalar field, coupled to the detector in the interaction Hamiltonian, is characterized by the field operator $\hat{\phi}(x)$. Expanding $\hat{\phi}(x)$ into its positive and negative frequency modes, we choose the planewave basis such that
\begin{align} \label{eq:field_operator}
    \hat{\phi}(x) = \int \frac{d^3k}{(2\pi)^{3/2}} \sqrt{\frac{\nu^2}{2 \omega(\bm{k})}} \Big ( \hat{a}(\bm{k}) e^{-i k \cdot x} + \mathrm{H.c.} \Big ) \, ,
\end{align}
where $\nu$ is the propagation speed of the field, and we have restricted our analysis to Minkowski spacetime. A full derivation of $\hat{\phi}(x)$, and summary of conventions employed, can be found in the Appendix. We distinguish the propagation speed of the field from the speed of light in our analysis for two reasons:
\begin{enumerate}[label=(\roman*)]
    \item to separate the relativistic nature of the detector from the field, which allows for easy determination of the detector's non-relativistic limit;
    \item for the study of detectors in media (which has been previously treated in~\cite{stritzelberger_coherent_2020, ginzburg_excitation_1986,*frolov_excitation_1986, brevik_quantum_1988,*brevik_quantum_1989}), and for possible connections to analogue models of relativity~\cite{todd_sound_2017, todd_particle_2021}.
\end{enumerate}
The standard treatment of a scalar field in a vacuum can be re-obtained by taking~$\nu = c$, which in natural units is~$\nu = 1$. The explicit form of the interaction Hamiltonian depends on the model of the detector and its coupling with the field. In the simplest case, one has a pointwise coupling between the field and monopole detector, but this is often extended to include a spatial smearing in the detector-field coupling, which we discuss in the following section.

\subsection{Point-like and smeared monopole detectors}
\label{sec:traditional_detectors}

The point-like detector model is generally treated as an idealization of a particle confined to a box, or alternatively as an atom, ion or molecule, which can be modeled as a two-level system with ground state $\ket{g}$ and excited state $\ket{e}$, and energies $E_g$ and $E_e$ respectively. Furthermore, a ``particle'' is said to be detected when the detector transitions from its ground state to its excited state. While the detector has quantized internal states, it is assumed to have a classical~\ac{com} and to follow a corresponding classical worldline~$\bm{(} t(\tau), \bm{x}(\tau) \bm{)}$, which is conventionally parametrized by its proper time~$\tau$.

From these assumptions, the free Hamiltonian for the detector is simply given by
\begin{align}\label{eq:restHD}
    \hat{H}_D = E \ket{e}\bra{e} \, ,
\end{align}
where $E = E_e - E_g$ denotes the energy gap, and without loss of generality we have taken $E_g = 0$.

The interaction between the detector and scalar field is most often treated as a linear coupling, taken as analogous to the atom-light interaction between a dipole and quantum electromagnetic field $-\hat{\bm{d}} \cdot \hat{\bm{E}}$. Such a form of the Hamiltonian for the detector-field interaction, which generates time-evolution with respect to the detector's proper time~$\tau$ in the interaction picture, is described by
\begin{align} \label{eq:pointlike_interaction}
    \hat{H}_I(\tau) = \lambda \hat{\mu}(\tau) \otimes \hat{\phi}\bm{(} x(\tau) \bm{)} \, ,
\end{align}
where $\lambda$ is the coupling strength, $\hat{\phi}$ is the aforementioned quantized scalar field, and $\hat{\mu}$ is the monopole operator for the detector, whose time-evolution is defined by
\begin{align}
    \hat{\mu}(\tau) = \ket{g}\bra{e} e^{-i E \tau} + \ket{e}\bra{g} e^{i E \tau} \, ,
\end{align}
and enables the field to excite and de-excite the detector.

The simple coupling between the detector and field in Eq.~\ref{eq:pointlike_interaction} is commonly considered in the literature, where the interaction is generally assumed to be time-independent. However, if the interaction is confined to a finite time interval, then one encounters ultraviolet divergences in the response of the detector~\cite{svaiter_inertial_1992}. These divergences can be removed by allowing the coupling to vary in time, such that~$\lambda \to \lambda(\tau)$, and requiring the interaction to be smoothly switched on and off~\cite{higuchi_uniformly_1993, sriramkumar_finite-time_1996, satz_then_2007}.

Further problems arise from the point-like nature of the detector model. Ironically, while the point-like approximation originally introduced by~\citet{dewitt_quantum_1979} was intended to simplify the model, it leads to divergences that must in turn be regularized by modeling the finite extension of the detector. As treated in~\cite{takagi_vacuum_1986, schlicht_considerations_2004, langlois_causal_2006, louko_how_2006}, these problems are resolved by modeling the spatial profile of the detector, which can be introduced by modifying the interaction Hamiltonian to include a smeared coupling between the detector and field, i.e.,~$\lambda \to \lambda(\bm{x})$.

To ensure that the response of the detector is safe from the above-mentioned problems, both the switching of the interaction and the finite extension of the detector must be included in the model. For simplicity, one may assume the interaction to be smeared over a spacelike hypersurface, here parametrized by the coordinate time~$t$. In this case, the time-dependent coupling with smearing can be decomposed as
\begin{align}
    \lambda(\bm{x}, t) = \lambda \chi(t) F(\bm{x} - \bm{x}_D) \, ,
\end{align}
where~$\chi(t)$ is the switching function, and~$F(\bm{x} - \bm{x}_D)$ is the smearing function centered around the stationary detector's~\ac{com} at~$\bm{x} = \bm{x}_D$. The interaction Hamiltonian is obtained by integrating over the spatial extent of the detector
\begin{align} \label{eq:pointlike_smearing}
    \hat{H}_I(t) = \lambda \chi(t) \int d^3x \, F(\bm{x} - \bm{x}_D) \hat{\mu}(t) \otimes \hat{\phi}(\bm{x}, t) \, ,
\end{align}
The switching and smearing functions are smooth functions, commonly chosen to have compact support, or to have a Gaussian or Lorentzian profile. While one largely has freedom to choose the switching function for the interaction, the introduction of a smearing profile is physically dubious if done in an \textit{ad hoc} manner.

Instead, one can derive the smearing profile from first-principles by considering a specific physical description for the detector. For example, the smearing associated with the discrete energy levels of a particle trapped in a box~\cite{unruh_what_1984}, or the smearing of a two-level atom given by its atomic wavefunction~\cite{martin-martinez_wavepacket_2013}.

While the introduction of a spatial smearing is sufficient for the problems caused from the point-like idealization, the model still assumes a detector following a classical trajectory. A fully quantum treatment would include the quantized degrees of freedom associated with the detector's \ac{com}. Moreover, by quantizing the \ac{com} degrees of freedom, the motivations for introducing a smearing profile do not arise; thus, smearing is not considered in the subsequent models.


\subsection{Detectors with quantized center of mass}
\label{sec:first_quantized_non_rel}

To provide a more physical account of the detector and its smearing, one can take the detector to be localized inside a box, rather than assume a point-like monopole. Such a model was given by~\citet{unruh_what_1984}, who considered a scalar field coupled to a detector with a quantized~\ac{com}, i.e.,~$\bm{x}_D \to \hat{\bm{x}}_D$ describes the position operator of the detector. The interaction between the box-detector and field in the Schr\"{o}dinger picture is
\begin{align}
    \hat{H}_I = \lambda \int d^3x \, \delta^{(3)}(\bm{x} - \hat{\bm{x}}_D) \hat{\phi}(\bm{x}) \, ,
\end{align}
and by introducing a completeness relation over position states, can equivalently be written as
\begin{align}
    \hat{H}_I = \lambda \int d^3x \, \ket{\bm{x}}\bra{\bm{x}}_D \otimes \hat{\phi}(\bm{x}) \, .
\end{align}
Because the detector is confined in a box, we may express the above in terms of the detector's energy eigenstates~$\ket{j}$. Further insertion of completeness relations gives
\begin{align}
    \hat{H}_I &= \lambda \int d^3x \, \sum_j \ket{j}\bra{j} \ket{\bm{x}}\bra{\bm{x}}_D \sum_k \ket{k}\bra{k} \otimes \hat{\phi}(\bm{x}) \notag \\
    &= \lambda \int d^3x \, \sum_{j, k} \psi_j^*(\bm{x}) \psi_k(\bm{x}) \ket{j}\bra{k} \otimes \hat{\phi}(\bm{x}) \, ,
\end{align}
where $\psi_j(\bm{x}) \equiv \braket{\bm{x} | j}$ denotes the position-space wavefunction for the $j$th energy level. Following~\cite{unruh_what_1984}, one may restrict the detector to two energy levels $j = \{ g, e \}$ and drop diagonal terms in the interaction, i.e.,~$\ket{g}\bra{g}$ and $\ket{e}\bra{e}$. One thereby obtains an interaction with a position-dependent monopole operator
\begin{align}
    \hat{H}_I = \lambda \int d^3x \, \hat{\mu}(\bm{x}) \otimes \hat{\phi}(\bm{x}) \, ,
\end{align}
where
\begin{align}
    \hat{\mu}(\bm{x}) \equiv \psi_g^*(\bm{x}) \psi_e(\bm{x}) \ket{g}\bra{e} + \psi_e^*(\bm{x}) \psi_g(\bm{x}) \ket{e}\bra{g} \, ,
\end{align}
which may be compared to the earlier introduction of smearing for the point-like detector~(\ref{eq:pointlike_smearing}).

A different extension of the \ac{udw} model, which idealizes the atom-light interaction, was considered by~\citet{stritzelberger_coherent_2020}. Their detector model extends the traditional \ac{udw} model to include the quantized \ac{com}, in addition to the usual quantized internal degrees of freedom for the two-level system. In their original paper, \nolink{\citeauthor{stritzelberger_coherent_2020}} consider an inertial detector whose evolution is described by the free non-relativistic Hamiltonian given by
\begin{align} \label{eq:first_q_free_hamiltonian}
    \hat{H}_D = \frac{\hat{\bm{p}}^2}{2 M} + E \ket{e}\bra{e} \, ,
\end{align}
which generates time-evolution with respect to the coordinate time $t$, and where $\hat{\bm{p}}$ is the~\ac{com} momentum operator and $M$ the detector's mass. As opposed to the traditional~\ac{udw} model, which features a detector following a classical trajectory, a~\ac{com}~rest frame (where $\bm{p} = \bm{0}$) cannot be freely chosen for a quantized~\ac{com} wavefunction, since this would violate the Heisenberg uncertainty relation. Therefore, the external~\ac{com} degrees of freedom must be included in the free evolution, where the detector states now live in the product Hilbert space~$\mathcal{H}_D~=~\mathcal{H}_{\mathrm{ext}}~\otimes~\mathcal{H}_{\mathrm{int}}$. The free Hamiltonian has eigenstates of the form
\begin{align}
    \hat{H}_D \ket{\bm{p}} \ket{j} = \left( \frac{\bm{p}^2}{2 M} + E_j \right) \ket{\bm{p}} \ket{j} \, ,
\end{align}
where $E_j$ is the $j$th energy level, corresponding to either the ground or excited state of the detector~$j = \{ g, e \}$.

The detector-field interaction has the same form as the box detector model of~\citet{unruh_what_1984}, but now features a monopole operator separate from the \ac{com} degrees of freedom, which couples to the field as
\begin{align} \label{eq:first_q_int_hamiltonian}
    \hat{H}_I &= \lambda \, \hat{\mu} \otimes \hat{\phi}(\hat{\bm{x}}) \notag \\
    &= \lambda \int d^3x \, \hat{\mu} \otimes \ket{\bm{x}}\bra{\bm{x}}_D \otimes \hat{\phi}(\bm{x}) \, ,
\end{align}
where the detector's monopole and~\ac{com} are evolved with respect to the detector's free Hamiltonian~(\ref{eq:first_q_free_hamiltonian})
\begin{subequations}
\begin{align}
    \hat{\mu}(t) &= e^{i \hat{H}_D t} \, \hat{\mu}(0) \, e^{-i \hat{H}_D t} \, , \\
    \ket{\bm{x}(t)}\bra{\bm{x}(t)}_D &= e^{i \hat{H}_D t} \ket{\bm{x}}\bra{\bm{x}}_D e^{-i \hat{H}_D t} \, .
\end{align}
\end{subequations}
Given the non-relativistic form of Eq.~\eqref{eq:first_q_free_hamiltonian}, the Hamiltonian is no longer Lorentz invariant and cannot describe relativistic trajectories. Moreover, as discussed in the Introduction, the mixing of the Galilean and Lorentz groups between the detector and field leads to non-physical effects, particularly in the full atom-light interaction, although this can be partially remedied by inclusion of relativistic corrections, such as the R\"{o}ntgen term.

In the case of the~\ac{udw} model proposed by~\citet{stritzelberger_coherent_2020}, an incorporation of the detector's mass defect was recently studied in~\cite{wood_quantized_2022}, whereby one models the change in the detector's mass-energy resulting from emission and absorption. Such a ``semi-relativistic'' approach is advantageous insofar that one wants to avoid a relativistic treatment, although a fully Lorentz covariant approach can be developed only in a relativistic second-quantized model.


\subsection{Second-quantized detector models}\label{sec:second_quantized}

In addition to the box detector model, \citet{unruh_notes_1976} proposed a fully relativistic second-quantized model, where the detector was represented by a composite scalar field. Despite its initial proposal almost a half-century ago, the second-quantized model has seen comparatively little study. A likely reason for this unpopularity is that one loses the convenient localization, and simplicity, of the point-like~\ac{udw} model. Despite these disadvantages, a full consideration of relativistic effects in a detector model is certainly of interest, and has been previously considered in a number of recent analyses~\cite{parentani_recoils_1995, casadio_accelerated_1995, giacomini_second-quantized_2022}.

In the past treatments, the second-quantized detector has generally been assumed to stay confined to some region, such as in a box-detector by \nolink{\citeauthor{unruh_notes_1976}} or the field in a cavity considered by~\citet{grove_notes_1983}. A disadvantage of these models is that additional care must be taken to define the rigidity of the container, namely how one should formalize the detectors' walls, but is generally resolved by taking the walls to remain fixed relative to some inertial observer.

However, one might well ask whether a full treatment of the detector's container is really necessary, particularly given the simplicity of the \ac{udw} model where no mention or treatment of a box is required. Indeed, due to the complexities of the second-quantized model, it is much easier to simplify the model to a monopole detector by restricting to the low-energy regime~\cite{grove_notes_1983}. Likewise, in a separate analysis by~\citet{colosi_what_2008}, found that for a box-detector sufficiently large with respect to its Compton wavelength~$\lambda_c \equiv \hbar / m c$, the correlation functions of the localized states exponentially converge to the global states defined on Minkowski spacetime~\cite{Note2}. Consequently, one can justify neglecting the detector's container on the basis that one remains in the low-energy regime for large wavelengths with respect to the size of the detector, and where boundary effects are neglected.

As in previous treatments~\cite{unruh_notes_1976, parentani_recoils_1995, casadio_accelerated_1995, giacomini_second-quantized_2022}, the detector is modeled as a composite real-valued scalar field~$\hat{\psi}_j$, for which we consider a continuum normalization. As in the first-quantized case, the detector has two energy levels $j = \{ g, \, e \}$ corresponding respectively to the ground and excited states. In the Schr\"{o}dinger picture, the detector is defined by
\begin{align} \label{eq:detector_second_q}
    \hat{\psi}_j(\bm{x}) &= \int \frac{d^3p}{(2\pi)^{3/2} \sqrt{2 E_j(\bm{p})}} \left( \hat{b}_j(\bm{p}) e^{i \bm{p} \cdot \bm{x}} + \mathrm{H.c.} \right) \, ,
\end{align}
with $E_j(\bm{p}) = \sqrt{\bm{p}^2 + M_j^2}$, and total mass-energy $M_j = m + E_j$ given by the sum of the detector's rest mass and internal energy. The creation and annihilation operators satisfy the non-covariant commutation relations $[\hat{b}_j(\bm{p}), \, \hat{b}^\dagger_k(\bm{p}')] = \delta^{(3)}(\bm{p} - \bm{p}') \delta_{jk}$.

The free Hamiltonian for the detector is given by the sum of two fields describing the internal energies
\begin{align} \label{eq:free_hamiltonian_second_q}
    \hat{H}_D &= \sum_j \int d^3p \, E_j(\bm{p}) \, \hat{b}^\dagger_j(\bm{p}) \hat{b}_j(\bm{p}) \, ,
\end{align}
while the Hamiltonian for the detector-field interaction is given by
\begin{align} \label{eq:int_hamiltonian_second_q}
    \hat{H}_I = \lambda^{(\mathrm{2nd})} \int d^3x \, \sum_{j \neq k} \hat{\psi}_j(\bm{x}) \hat{\psi}_k(\bm{x}) \otimes \hat{\phi}(\bm{x}) \, .
\end{align}
In~\cite{giacomini_second-quantized_2022}, to simplify the model's structure, the interaction Hamiltonian was restricted to the single-particle sector, resulting in 
\begin{align} \label{eq:int_hamiltonian_second_q_one_sector_final}
    \hat{H}^{(1)}_I = \lambda^{(\mathrm{2nd})} \int d^3x \, \sum_{j \neq k} \ket{\bm{x}_j}\bra{\bm{x}_k}_D^{(\mathrm{2nd})} \otimes \hat{\phi}(\bm{x}) \,
\end{align}
which is close, but not identical to, the first-quantized interaction \eqref{eq:first_q_int_hamiltonian} introduced in~\cite{stritzelberger_coherent_2020}. Before we discuss the above interaction, it is worth noting that the restriction to the one-particle sector is in some sense redundant. Applying this interaction to an initial state containing, say, one detector yields final states with also just one detector, possibly with a different internal energy. In this respect, Eq.~\eqref{eq:int_hamiltonian_second_q} preserves the total number of detectors~$\hat{\psi}$.


In addition to the different dispersion relations for the~\ac{com} of the detector between Eqs.~\eqref{eq:first_q_int_hamiltonian} and~\eqref{eq:int_hamiltonian_second_q_one_sector_final}, one finds that the second-quantized model has non-orthogonal ``position states'' defined by
\begin{align} \label{eq:second_quantized_position_state}
    \ket{\bm{x}_j}_D^{(\mathrm{2nd})} &\equiv \int \frac{d^3p}{(2\pi)^{3/2} \sqrt{2 E_j(\bm{p})}} e^{-i \bm{p} \cdot \bm{x}} \hat{b}_j^\dagger(\bm{p}) \ket{0}_D \, ,
\end{align}
which are equivalently given by
\begin{align}
    \ket{\bm{x}_j}_D^{(\mathrm{2nd})} &= \hat{\psi}_j(\bm{x}) \ket{0}_D \, .
\end{align}
According to the common textbook interpretation, the above states constitute position states of the second-quantized theory (see, for example, Eq.~(2.41) and successive comments in~\cite{peskin_introduction_2019}); however, the presence of the $1 / \sqrt{2 E_j(\bm{p})}$~integration measure obfuscates such a straightforward view. Clearly, while these second-quantized ``position states'' transform covariantly, they do not correspond to eigenstates of a position operator, or any Hermitian operator for that matter, a fact which was pointed out in~\cite{barros_e_sa_quantum_2021}.


The aforementioned non-orthogonality of these ``position states'' is also due to the presence of the integration measure. Indeed, these states do not correspond to a Fourier transform in momentum space, as is the case for the first-quantized position states~\cite{Note3}. This non-orthogonality is generally interpreted as a non-localizability in \ac{rqm}, which has been the subject of extensive discussion in the literature, and has recently been studied in several papers analyzing the transition from relativistic \ac{qft} to \ac{nrqm}~\cite{padmanabhan_obtaining_2018, papageorgiou_impact_2019, barros_e_sa_quantum_2021}.

To summarize, one can easily define a second-quantized model of the detector dynamics, which in the single-particle sector has an interaction that resembles, but is not identical to, the first-quantized treatment due to the different localization that one obtains. However, if this resulting second-quantized model is adopted, then the question arises whether one can start directly from a relativistic first-quantized treatment of the detector. Such a model would already be constrained to the single-particle sector and would necessarily treat the detector as a localized system. In the following section, we therefore construct and analyze such a first-quantized model from the standpoint of \ac{rqm}, and compare its predictions to both the relativistic second-quantized model, as well as the semi- and non-relativistic regimes.


\section{Relativistic detector with first-quantized center of mass}\label{sec:rel_first_q_model}

In our review of past detector models, the physical description of the detector is often abstracted in order to focus exclusively on the field. Generally, in the context of quantum field theory, particle detectors are introduced in order to give an operational meaning to particles, i.e.,~particles are what particle detectors detect. There are a number of remarks to be made here, the first being that the idealization of a detector by the monopole model leads to problems which must be resolved by reintroducing the very structure that was originally abstracted away. Second, the operational definition of a particle detector risks circularity; just as a ``particle'' has some effective ontology, so too does a particle detector. Care must be taken to adequately define and formalize what precisely one means by a ``detector.''

Here, we consider a detector to be a suitably localized system such that one can employ a first-quantized formulation, but where the free dynamics \textit{may not} be abstracted. In formulating a relativistic first-quantized model, we take the interaction to be of the same general form as in Eqs.~\eqref{eq:first_q_int_hamiltonian} and~\eqref{eq:int_hamiltonian_second_q_one_sector_final}
\begin{align}
    \hat{H}_I &= \lambda^{(\mathrm{1st})} \int d^3x \, \hat{\mu} \otimes \ket{\bm{x}}\bra{\bm{x}}_D^{(\mathrm{1st})} \otimes \hat{\phi}(\bm{x}) \notag \\
    &= \lambda^{(\mathrm{1st})} \int d^3x \, \sum_{j \neq k} \ket{\bm{x}, j}\bra{\bm{x}, k}_D^{(\mathrm{1st})} \otimes \hat{\phi}(\bm{x}) \, ,
\end{align}
where we have combined the detector's external and internal states into the first-quantized position states. Note that we distinguish between the first- and second-quantized coupling constants,~$\lambda^{(\mathrm{1st})}$ and~$\lambda^{(\mathrm{2nd})}$, which have different dimensions between the two models. We take the first-quantized position states to be equivalent to the non-relativistic theory, defined by
\begin{align} \label{eq:first_quantized_position_state}
    \ket{\bm{x}}_D^{(\mathrm{1st})} \equiv \int \frac{d^3p}{(2\pi)^{3/2}} e^{-i \bm{p} \cdot \bm{x}} \ket{\bm{p}}_D \, .
\end{align}
These first-quantized position states, unlike those derived from the second-quantized theory, do not transform covariantly~\cite{Note4}, but are orthogonal and correspond to eigenstates of a respective position operator. More specifically, they correspond to the Newton-Wigner position operator~\cite{newton_localized_1949}, which was derived by invariance conditions that a suitable position operator should reasonably satisfy. Later, in Sec.~\ref{sec:discussion}, we shall discuss the two localizations in more detail; at this stage 
we concurrently consider both the first- and second-quantized 
models and subsequently compare their predictions. 

Working in the interaction picture, we evolve the position states with respect to the time coordinate~$t$
\begin{align} \label{eq:rqm_interaction}
    \hat{H}_I(t) &= \lambda \int d^3x \, \sum_{j \neq k} \ket{\bm{x}(t), j}\bra{\bm{x}(t), k}_D \otimes \hat{\phi}(\bm{x}, t) \, ,
\end{align}
with the time-evolutions for the first- and second-quantized position states respectively defined by
\begin{subequations}
\begin{align}
    \ket{\bm{x}(t), j}_D^{(\mathrm{1st})} &\equiv e^{i \hat{H}_D t} \ket{\bm{x}, j}_D^{(\mathrm{1st})} \, , \\
    \ket{x_j}_D^{(\mathrm{2nd})} &\equiv \hat{\psi}_j(x) \ket{0}_D \, ,
\end{align}
\end{subequations}
and the detector in the Heisenberg picture is given by
\begin{align}
    \hat{\psi}_j(x) = \int \frac{d^3p}{(2\pi)^{3/2} \sqrt{2 E_j(\bm{p})}} \left( \hat{b}_j(\bm{p}) e^{i p \cdot x} + \mathrm{H.c.} \right) \, .
\end{align}
For convenience, we introduce the general position state
\begin{align}
    \ket{\bm{x}, j}_D \equiv \int \frac{d^3p}{(2\pi)^{3/2} f_j(\bm{p})} e^{-i \bm{p} \cdot \bm{x}} \ket{\bm{p}}_D \, ,
\end{align}
where $f_j(\bm{p})$ is the integration measure for the second- ($f_j(\bm{p}) \equiv \sqrt{2 E_j(\bm{p})}$) and first-quantized ($f_j(\bm{p}) \equiv 1$) position states, cf.~Eqs.~(\ref{eq:second_quantized_position_state}) and (\ref{eq:first_quantized_position_state}) respectively.

Considering the detector as a classical system with mass~$M$, the rest frame Hamiltonian would be given by
\begin{align*}
    H_D^\tau = M \, ,
\end{align*}
where we explicitly denote that time-evolution is generated with respect to the detector's proper time~$\tau$. Rewriting this in terms of the coordinate time~$t$, we obtain
\begin{align*}
    H_D^t = \frac{d\tau}{dt} M \, ,
\end{align*}
acquiring a Lorentz factor from time-reparameterization of the equations of motion~\cite{martin-martinez_relativistic_2018}. While we restrict our analysis to Minkowski spacetime, one can in principle extend this treatment to curved spacetimes~\cite{martin-martinez_general_2020, perche_localized_2022}. Expressing the free dynamics in momentum space, and quantizing in this frame, one would conventionally consider a Hamiltonian of the form
\begin{align*}
    \hat{H}_D^t = \sqrt{\hat{\bm{p}}^2 + M^2} \, ,
\end{align*}
where the mass is treated as a $c$-number. However, this expression would be inappropriate for this model since the above Hamiltonian is unable to characterize the detector's internal degrees of freedom. Instead, we include a quantized mass-energy operator~\cite{zych_quantum_2011, pikovski_universal_2015, smith_quantum_2020, zych_quantum_2018}
\begin{align} \label{eq:rqm_free}
    \hat{H}_D^t = \sqrt{\hat{\bm{p}}^2 + \hat{M}^2} \, ,
\end{align}
 whose eigenvalues are given by the action on a corresponding mass eigenstate
\begin{align}
    \hat{M} \ket{j} = M_j \ket{j} \, .
\end{align}
For a detector modeled as a two-level system, one may express the mass-energy operator by
\begin{align} \label{eq:mass_energy_operator_two_level}
    \hat{M} = M_g \ket{g}\bra{g} + M_e \ket{e}\bra{e} \, ,
\end{align}
with~$M_j = m + E_j$ corresponding to the total mass-energy of the detector, as in the second-quantized model. Note that such a model is consistent with the conventional~\ac{udw} Hamiltonian,~i.e.,~Eq.~\eqref{eq:restHD} is recovered up to a dynamically irrelevant constant (the ground state mass-energy of the detector) for states with vanishing momentum. Aside from the introduction of a mass operator in the model just outlined, there are various other motivations for quantizing the mass-energy in~\ac{rqm}, which we outline and discuss in Sec.~\ref{sec:discussion}.

\subsection{Calculation of spontaneous emission rate}

For simplicity, we consider only a single physical process, namely spontaneous emission, which is primarily done to more easily compare with past analyses~\cite{stritzelberger_coherent_2020, wood_quantized_2022}, and because it is a frequent subject of study in the~\ac{udw} model, particularly from the context of quantum optics. One can also consider alternate physical processes, such as absorption and vacuum excitation, or in principle obtain more general results by formalizing in terms of correlations functions, which we defer to future work.

In the case of spontaneous emission, one starts with an excited detector in the vacuum
\begin{align} \label{eq:initial_state_em}
    \ket{\Psi_i} = \ket{\psi_i, e}_D \otimes \ket{0} \, ,
\end{align}
with an arbitrary center-of-mass wavefunction $\psi_i$ of the detector, which can be represented in the momentum basis as
\begin{align}
    \ket{\psi_i} = \int{d^3p_i \, \psi_i(\bm{p}_i) \ket{\bm{p}_i}} \, .
\end{align}
For the final state, the detector is in its ground state with a spontaneously emitted field quantum
\begin{align}
    \ket{\Psi_f} = \ket{\bm{p}_f, g}_D \otimes \hat{a}^\dagger(\bm{k}) \ket{0} \, .
\end{align}
Expanding the evolution operator perturbatively up to first order, we obtain
\begin{align} \label{eq:dyson_series_perturbation}
    \hat{U}(t_f, \, t_i) = \hat{\mathbbm{1}} - i \int_{t_i}^{t_f} dt \, \hat{H}_{I}(t) + \mathcal{O}(\lambda^2) \, .
\end{align}
The resulting transition amplitude evaluates to
\begin{multline} \label{eq:trans_amp}
    \mathcal{A}_{\ket{\bm{p}_i, e, 0} \to \ket{\bm{p}_f, g, 1_{\bm{k}}}} = \frac{-i \lambda}{(2\pi)^{3/2}} \sqrt{\frac{\nu^2}{2 \omega(\bm{k})}} \frac{\psi_i(\bm{p}_f + \bm{k}) }{f_g(\bm{p}_f) f_e(\bm{p}_f + \bm{k})} \\
    \times \int_{t_i}^{t_f} dt \, e^{i \left( E_g(\bm{p}_f) - E_e(\bm{p}_f + \bm{k}) + \omega(\bm{k}) \right) t} \, .
\end{multline}
It is most interesting to find the total transition rate, which can be obtained by first deriving the total probability of the detector's final internal state, given by
\begin{align}
    P_{\ket{\bm{p}_i, e, 0} \to \ket{g}} &= \int d^3k \int d^3p_f \, \left| \mathcal{A}_{\ket{\bm{p}_i, e, 0} \to \ket{\bm{p}_f, g, 1_{\bm{k}}}} \right|^2 \, ,
\end{align}
where we have traced over the final momenta for the field and detector.

In conventional analyses of the~\ac{udw} model, one separates the detector's selectivity from the response function of the field and only considers the latter~\cite{birrell_quantum_1982}. However, due to the modified form of both the interaction~(\ref{eq:rqm_interaction}) and free Hamiltonian~(\ref{eq:rqm_free}), the internal and external degrees of freedom of the detector are coupled, and as a result its dynamics can no longer be excluded. Therefore, we consider the full transition rate rather than just the response rate of the field. Substituting the transition amplitude~(\ref{eq:trans_amp}), one finds
\begin{multline}
    P[\psi_i] = \frac{\lambda^2}{(2\pi)^{3}} \int d^3k \int d^3p_f \, \frac{\nu^2}{2 \omega(\bm{k})} \frac{|\psi_i(\bm{p}_f + \bm{k})|^2}{f_g^2(\bm{p}_f) f_e^2(\bm{p}_f + \bm{k})} \\
    \times \int_{t_i}^{t_f} \int_{t_i}^{t_f} dt \, dt' \, e^{i \left( E_g(\bm{p}_f) - E_e(\bm{p}_f + \bm{k}) + \omega(\bm{k}) \right) (t - t')} \, ,
\end{multline}
and recognizing that the above integrals over time depend only on the interval $s = t - t'$, but not on the total time $s' = t + t'$, we may find the transition rate $\dot{P}[\psi_i] \equiv (d/dt_f) P[\psi_i]$ over the interval $\Delta t = t_f - t_i$, given by
\begin{multline}
    \dot{P}[\psi_i] = \frac{\lambda^2}{(2\pi)^{3}} \int d^3k \int d^3p_f \, \frac{\nu^2}{2 \omega(\bm{k})} \frac{|\psi_i(\bm{p}_f + \bm{k})|^2}{f_g^2(\bm{p}_f) f_e^2(\bm{p}_f + \bm{k})} \\
    \times \int_{-\Delta t}^{\Delta t} ds \, e^{i \left( E_g(\bm{p}_f) - E_e(\bm{p}_f + \bm{k}) + \omega(\bm{k}) \right) s} \, .
\end{multline}
Since switching effects are neglected, we restrict to asymptotic times $\Delta t \to \infty$, and re-introducing the integration over the initial momentum $\bm{p}$, the above expression becomes
\begin{multline}
    \dot{P}[\psi_i] = \frac{\lambda^2}{(2\pi)^{2}} \int \frac{d^3p}{f_e^2(\bm{p})} \, |\psi_i(\bm{p})|^2 \int \frac{d^3k}{f_g^2(\bm{k} - \bm{p})} \, \frac{\nu^2}{2 \omega(\bm{k})} \\
    \times \delta\Big[ E_g(\bm{k} - \bm{p}) - E_e(\bm{p}) + \omega(\bm{k}) \Big] \, .
\end{multline}
At this point, we consider the different localizations for the first- and second-quantized models separately. For the case of a massless scalar field, the dispersion relation is $\omega(\bm{k}) = \nu |\bm{k}|$, and the transition rate evaluates to
\begin{align} \label{eq:transrate_final}
    \dot{P}[\psi_i] = \frac{\lambda^2}{2 \pi} \int d^3p \, |\psi_i(\bm{p})|^2 \, \mathcal{T}_{\mathrm{rel}}(\bm{p}) \, ,
\end{align}
where $\mathcal{T}$ denotes the ``template function,'' originally coined and defined for the non-relativistic case in~\cite{stritzelberger_coherent_2020}. Considering each localization separately, the relativistic template functions for the first- and second-quantized models are given by
\begin{widetext}
\begin{subequations} \label{eq:rel_template_function_compact}
\begin{align}
    \mathcal{T}_{\mathrm{rel}}^{\mathrm{(1st)}}(\bm{p}) &\equiv \frac{\nu}{\left( 1 - \nu^2 \right)^2} \left[ \left( 1 + \nu^2 \right) \sqrt{\bm{p}^2 + M_e^2} - \left( \frac{1}{|\bm{p}|} \sqrt{\bm{p}^2 + M_e^2} + \nu \right) \ell\left( \nu |\bm{p}|, \, \sqrt{\bm{p}^2 + M_e^2}, \, M_g \sqrt{\nu^2 - 1} \right) \right] \, ,  \label{eq:rel_1st_template_function_compact}
\end{align}
\begin{align}
    \mathcal{T}_{\mathrm{rel}}^{\mathrm{(2nd)}}(\bm{p}) &\equiv \frac{\nu \left[ |\bm{p}| - \ell\left( \nu |\bm{p}|, \, \sqrt{\bm{p}^2 + M_e^2}, \, M_g \sqrt{\nu^2 - 1} \right) \right]}{4 \left( 1 - \nu^2 \right) |\bm{p}| \sqrt{\bm{p}^2 + M_e^2}} \, ,  \label{eq:rel_2nd_template_function_compact}
\end{align}
\end{subequations}
\end{widetext}
where $\ell$ is an auxiliary function of the form~\cite{[{This auxiliary function appears very frequently in the context of integrals involving Bessel functions, see, e.g.,~\S 6.5-6.7 of~}]gradshteyn_table_2014}
\begin{align}
    \ell(a, \, b, \, c) \equiv \frac{1}{2} \left( \sqrt{\left( a + b \right)^2 + c^2} - \sqrt{\left( a - b \right)^2 + c^2} \, \right) \, .
\end{align}
The above expressions for an interaction in media 
simplify dramatically in the vacuum case, i.e.,~when taking $\nu = c = 1$. While it may initially appear that the template functions~(\ref{eq:rel_template_function_compact}) diverge, the results are well-defined in the limit $\nu \to 1$. The vacuum template function corresponding to the first-quantized localization is given by
\begin{subequations}
\begin{align} \label{eq:first_quantized_template_function_vacuum}
    \lim_{\nu \to 1} \mathcal{T}_{\mathrm{rel}}^{\mathrm{(1st)}}(\bm{p}) &= \frac{1}{4} \left( 1 - \frac{M_g^4}{M_e^4} \right) \sqrt{\bm{p}^2 + M_e^2} \, ,
\end{align}
while for the second-quantized localization one finds
\begin{align} \label{eq:second_quantized_template_function_vacuum}
    \lim_{\nu \to 1} \mathcal{T}_{\mathrm{rel}}^{\mathrm{(2nd)}}(\bm{p}) &= \frac{1}{8} \left( 1 - \frac{M_g^2}{M_e^2} \right) \frac{1}{\sqrt{\bm{p}^2 + M_e^2}} \, ,
\end{align}
\end{subequations}
where notably the second-quantized model features the reciprocal of the relativistic dispersion relation, in contrast to the first-quantized case; see also Fig.~\ref{fig:template_comparison_vac} for a comparison of these two localizations.

The different behavior between the two models results from the distinct integration measures, which enter in the first- and second-quantized position states, and is ultimately due to the different notions of ``local interaction'' between the field and detector in the two cases. Moreover, the different dimensions of the two template functions is offset by the effective difference in dimensions between the coupling constants of each model, which give the correct dimension for the transition rate.

Given that the coupling constants between the first- and second-quantized localizations are distinct, it is necessary to find a means to equate them so that the two cases can be meaningfully compared. Requiring that the first- and second-quantized template functions agree for~$|\bm{p}| = 0$, one obtains a mass-dependent factor relating the respective coupling constants by
\begin{align} \label{eq:coupling_relation}
    \lambda^{\mathrm{(2nd)}} = \sqrt{2 (M_g^2 + M_e^2)} \, \lambda^{\mathrm{(1st)}} \, .
\end{align}
Employing this relation between coupling constants, one obtains corrected expressions for Eqs.~\eqref{eq:rel_2nd_template_function_compact} and~\eqref{eq:second_quantized_template_function_vacuum}
\begin{widetext}
\begin{subequations} \label{eq:second_quantized_template_function_corrected}
\begin{align}
    \mathcal{T}_{\mathrm{rel}}^{\mathrm{(2nd)}}(\bm{p}) &\equiv \frac{\nu (M_g^2 + M_e^2) \left[ |\bm{p}| - \ell\left( \nu |\bm{p}|, \, \sqrt{\bm{p}^2 + M_e^2}, \, M_g \sqrt{\nu^2 - 1} \right) \right]}{2 \left( 1 - \nu^2 \right) |\bm{p}| \sqrt{\bm{p}^2 + M_e^2}} \, , \label{eq:second_quantized_template_function_medium_corrected}
\end{align}
\begin{align}
    \lim_{\nu \to 1} \mathcal{T}_{\mathrm{rel}}^{\mathrm{(2nd)}}(\bm{p}) &= \frac{1}{4} \left( 1 - \frac{M_g^4}{M_e^4} \right) \frac{M_e^2}{\sqrt{\bm{p}^2 + M_e^2}} \, , \label{eq:second_quantized_template_function_vacuum_corrected}
\end{align}
\end{subequations}
\end{widetext}
which agree in dimension with the first-quantized case. In all instances where we compare the first- and second-quantized localizations, such as for the template functions given in Figs.~\ref{fig:template_comparison_vac} and~\ref{fig:template_comparison_med}, we use the above results and in general employ Eq.~(\ref{eq:coupling_relation}).

\newpage \subsection{\texorpdfstring{Re-derivation of transition rates for\\
semi- and non-relativistic detectors}{Re-derivation of transition rates for semi- and non-relativistic detectors}}
\begin{figure*}
    \centering
    \subfloat[][$E / m = 0.001$]{
    \includegraphics[width=0.375\textwidth]{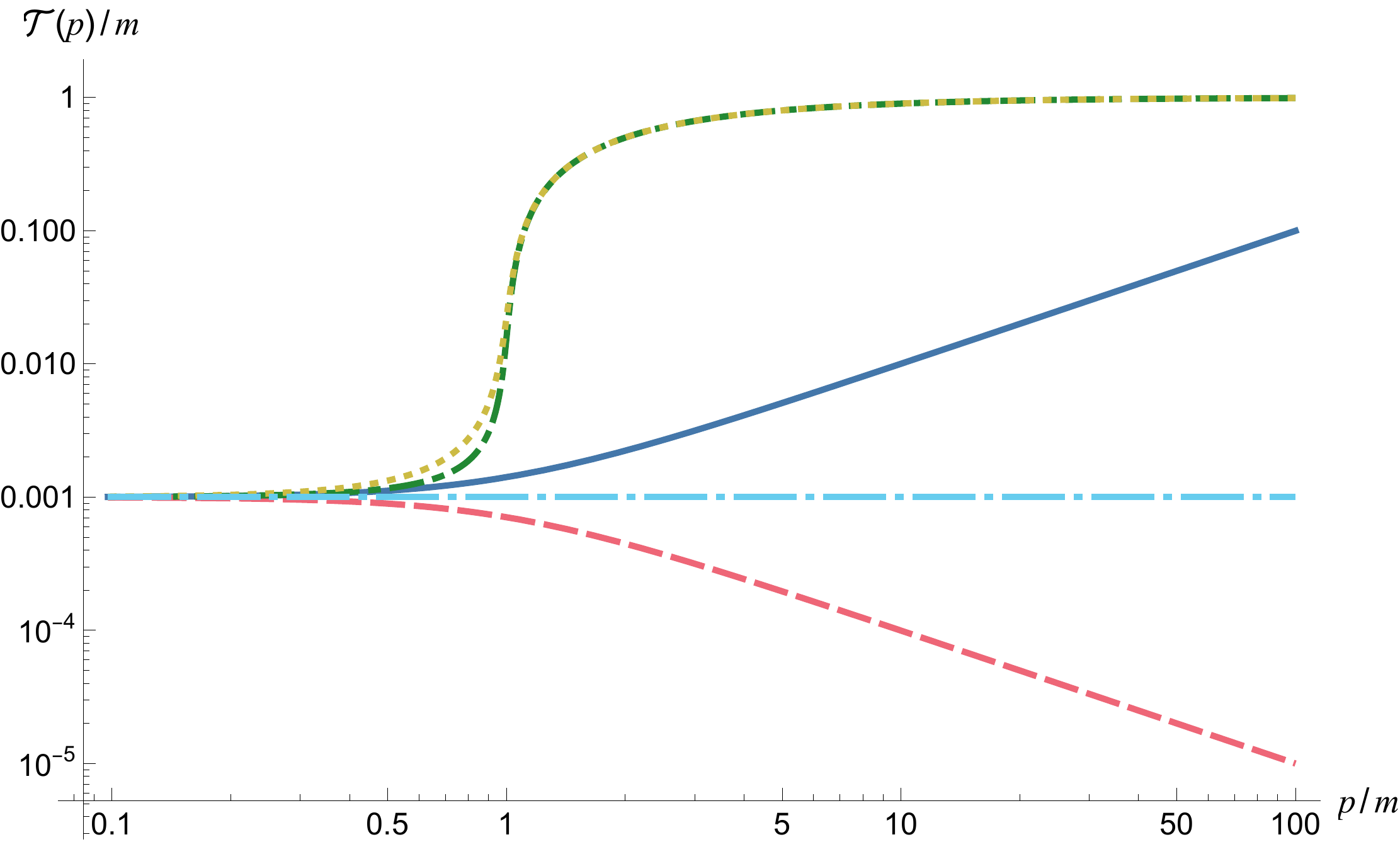}
    \label{fig:template_comparison_vac_E0001}}
    \qquad
    \subfloat[][$E / m = 10$]{
    \includegraphics[width=0.375\textwidth]{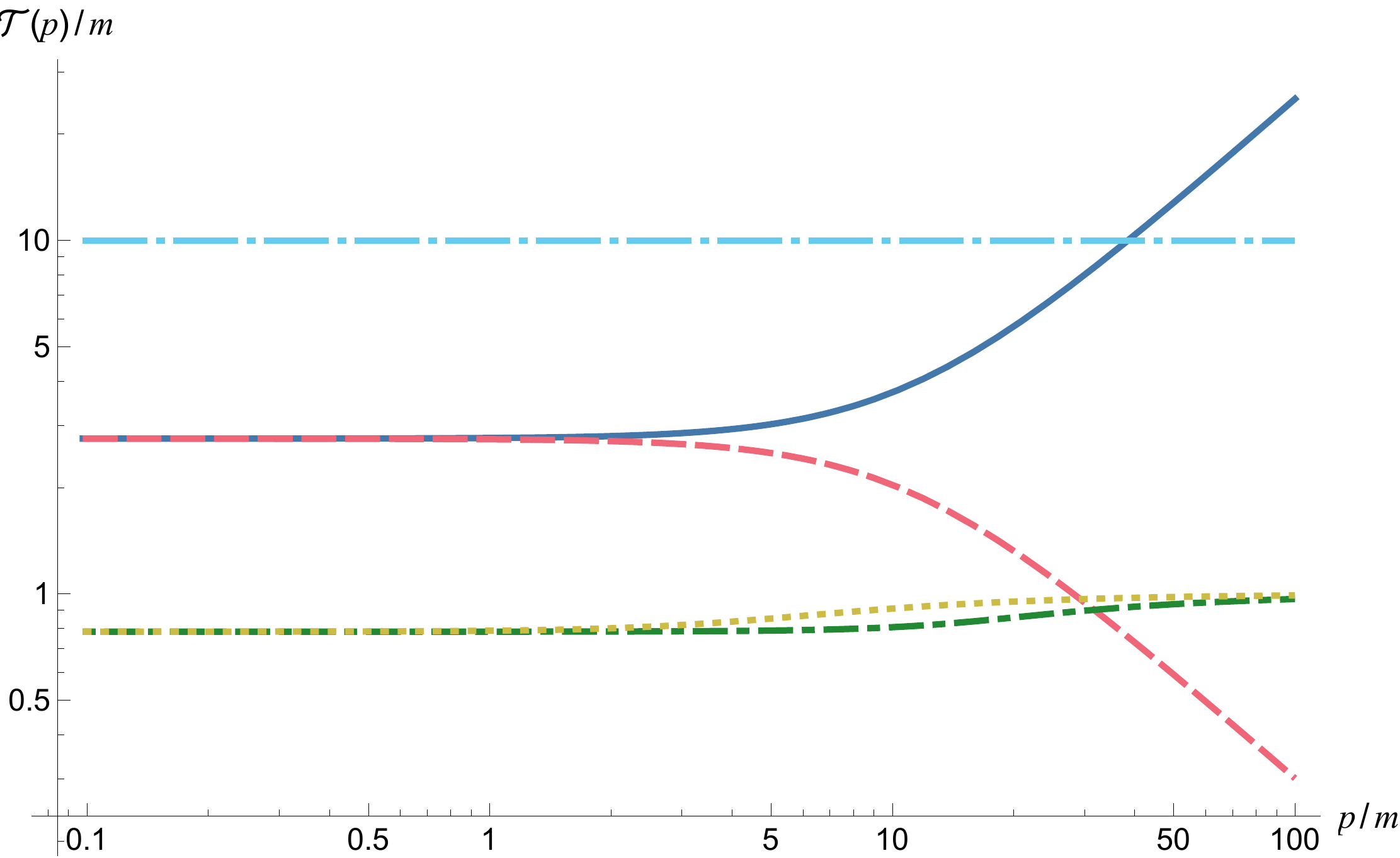}
    \label{fig:template_comparison_vac_E10}}
    \qquad
    \subfloat{\raisebox{12.5mm}%
    {\includegraphics[width=0.125\textwidth]{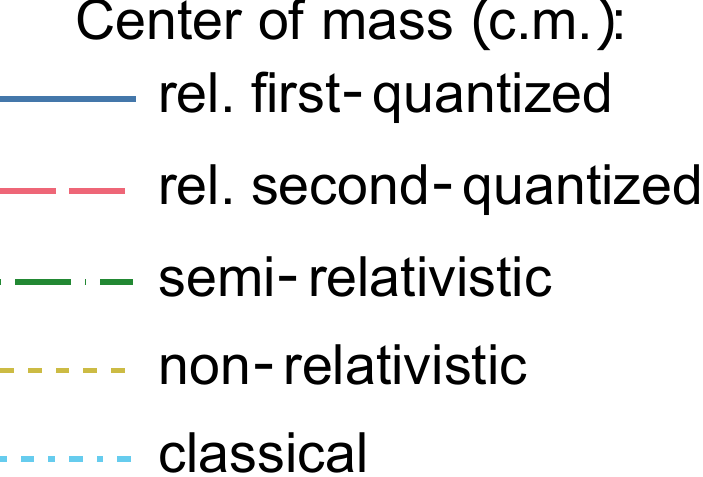}}}
    \caption{Template functions for the first- and second-quantized localizations in a vacuum~($\nu = c = 1$), alongside their semi- and non-relativistic limits. Results are given with respect to the Compton scale of the detector, where~$m$ is the detector's rest mass. For small energy gap (a), all cases coincide for small~$p / m$ (i.e.,~for small momenta or sufficiently massive detectors); for large energy gap (b), results no longer agree between the relativistic, classical and semi- and non-relativistic~\ac{com}, even for small~$p / m$. The distinction between the different template functions suggests that these cases can be empirically distinguished.}
    \label{fig:template_comparison_vac}
\end{figure*}
\begin{figure*}
    \centering
    \subfloat[][$\nu = 0.1$]{
    \includegraphics[width=0.375\textwidth]{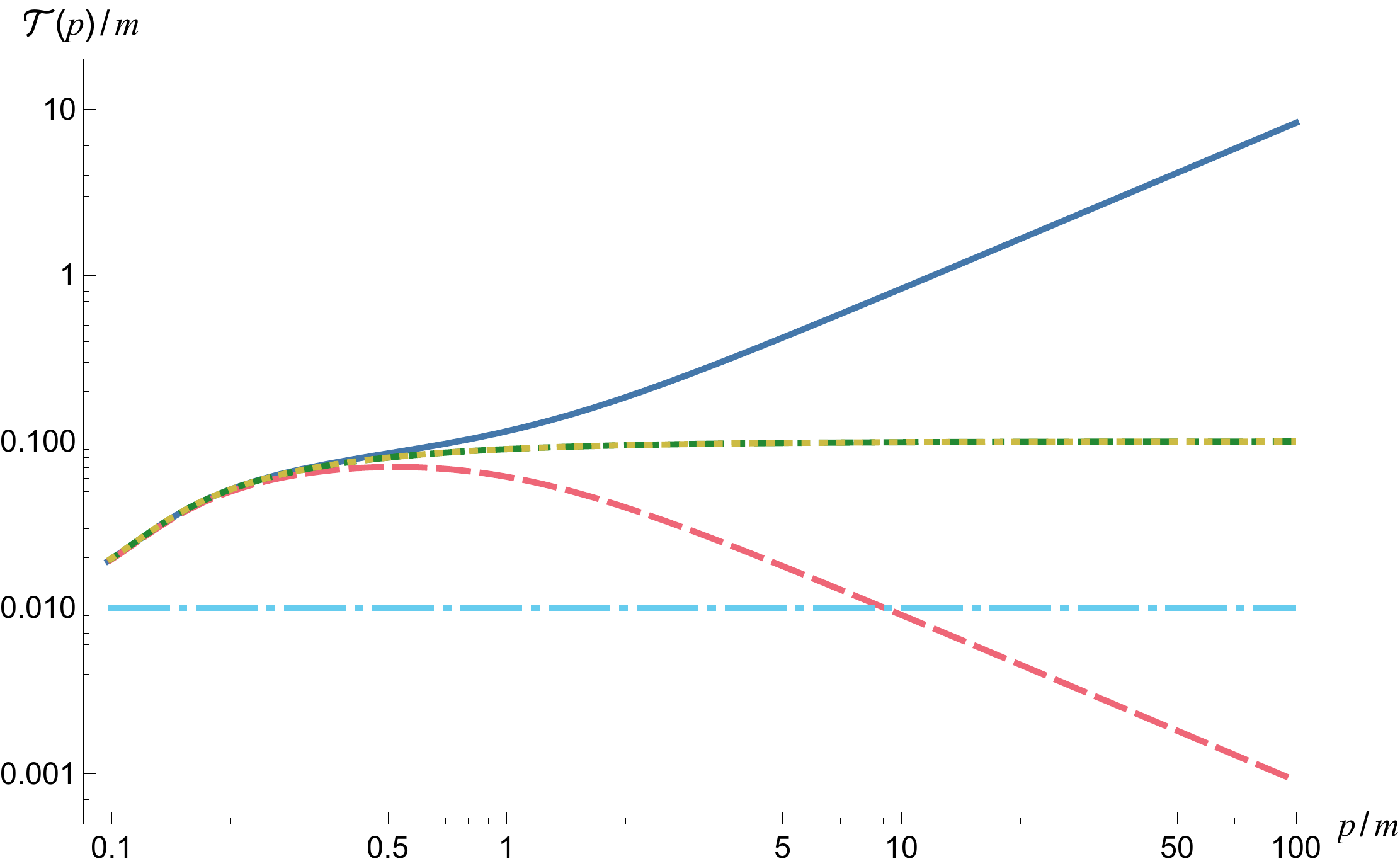}
    \label{fig:template_comparison_med_nu01}}
    \qquad
    \subfloat[][$\nu = 0.9$]{
    \includegraphics[width=0.375\textwidth]{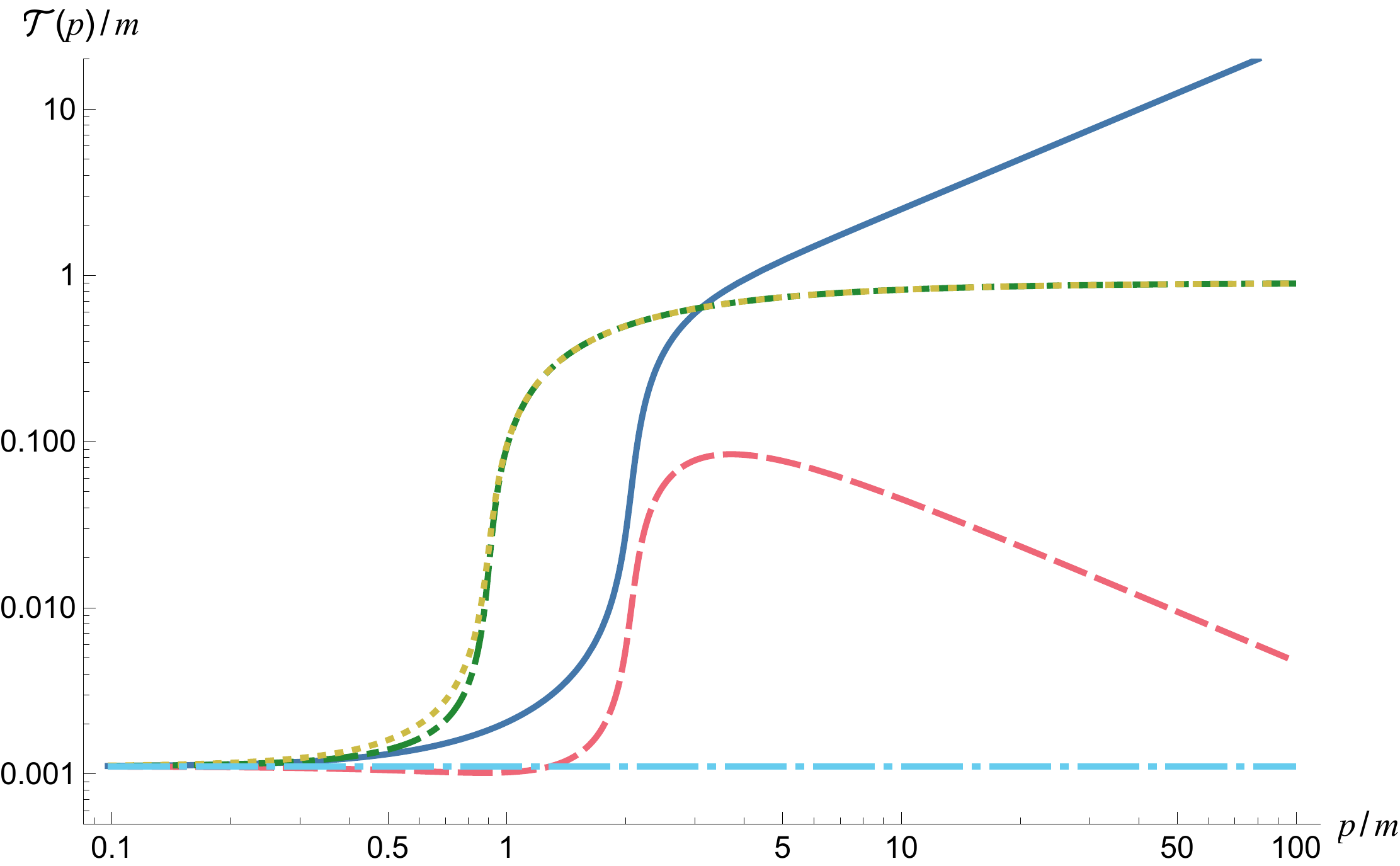}
    \label{fig:template_comparison_med_nu09}}
    \qquad
    \subfloat{\raisebox{12.5mm}%
    {\includegraphics[width=0.125\textwidth]{figures/template_comparison_legend.pdf}}}
    \caption{Template functions for energy gap~$E / m = 0.001$ between the first- and second-quantized localizations in a medium, alongside their semi- and non-relativistic limits. Results are given with respect to the Compton scale of the detector, where~$m$ is the detector's rest mass. In comparison to the vacuum case, one observes new transient behavior for small~$p / m$; in (a) the relativistic template functions closely follow their semi- and non-relativistic limits, while in (b) the relativistic template functions initially peak, reaching a local maximum dependent on the propagation speed of the field.}
    \label{fig:template_comparison_med}
\end{figure*}
While the template functions are particularly simple in the vacuum case, one is unable to recover exact analytical expressions found in the Galilean limit, where~$|\bm{p}| \ll M_j$. This is due to the fact that only the dynamics of the detector are governed by the Galilean group, while the field's dynamics are still described by the Lorentz group. Therefore, it is necessary to distinguish between the speed of light present in the detector's relativistic dispersion relation and the field's propagation speed, so that one has respectively separated the group structure governing the detector and field.

Consequently, one may take the ``limit'' of the Lorentz group to the Galilean group for the detector by subtracting the rest energy and taking~$|\bm{p}| \ll M_j$; by ``limit'' here, one formally means a contraction of a given Lie group, such that a new group structure is obtained from the limiting case of a parameter in the Lie algebra~\cite{inonu_contraction_1953}. Recall from Eq.~(\ref{eq:mass_energy_operator_two_level}) that the mass-energy for the two levels is defined by~$M_j = m + E_j$, where~$m\equiv M_g$ is the rest mass-energy of the detector for~$E_g = 0$. Contracting the free Hamiltonian~(\ref{eq:rqm_free}), one obtains
\begin{align}\label{eq:semi-rel-Hamilt}
    \sqrt{\hat{\bm{p}}^2 + \hat{M}^2} - m \hat{\mathbbm{1}} \to \frac{1}{2} \, \hat{\bm{p}}^2 \hat{M}^{-1} + E\ket{e}\bra{e} \, ,
\end{align}
where $E$ is the detector's internal energy gap (equivalently taking $E = E_e$). Notably, Eq.~\eqref{eq:semi-rel-Hamilt} still features a quantized mass-energy, as one has contracted to the centrally extended Galilean group in this limit. Regarding the template functions, the first-quantized case~(\ref{eq:rel_1st_template_function_compact}) reduces to
\begin{widetext}
\begin{align} \label{eq:semirel_1st_template_function_compact}
    \mathcal{T}_{\mathrm{semi-rel}}^{\mathrm{(1st)}}(\bm{p}) &= \frac{M_g \nu}{|\bm{p}|} \left[ |\bm{p}| - \ell\left( |\bm{p}|, \, M_g \nu, \, \sqrt{2 M_g E - p^2 \left( 1 - \frac{M_g}{M_e} \right)} \, \right) \right] \, ,
\end{align}
\end{widetext}
which was previously obtained in~\cite{wood_quantized_2022} directly using the ``semi-relativistic'' Hamiltonian~\eqref{eq:semi-rel-Hamilt} for the detector's~\ac{com}. In this semi-relativistic template function~(\ref{eq:semirel_1st_template_function_compact}), the parameter~$\nu$ has been previously interpreted as the speed of light, but should more precisely be understood as the propagation speed of the field, i.e.,~a parameter formally independent of the speed of light.


One can fully contract to the non-relativistic limit by taking $\hat{M} \to m \hat{\mathbbm{1}}$ with $M_g = M_e = m$, whereby the template function reduces to
\begin{align} \label{eq:nonrel_1st_template_function_compact}
    \mathcal{T}_{\mathrm{non-rel}}^{\mathrm{(1st)}}(\bm{p}) &= \frac{m \nu}{|\bm{p}|} \left[ |\bm{p}| - \ell\left( |\bm{p}|, \, m \nu, \, \sqrt{2 m E} \, \right) \right] \, ,
\end{align}
and one obtains the non-relativistic model analyzed in~\cite{stritzelberger_coherent_2020}. Comparing the semi- and non-relativistic models alongside the relativistic cases, as seen in Fig.~\ref{fig:template_comparison_vac}, the difference between the semi- and non-relativistic cases is small, but becomes particularly notable when comparing to the relativistic cases. As can be most easily seen in Fig.~\ref{fig:template_comparison_vac_E0001}, the relativistic first-quantized template function increases without bound, while the semi- and non-relativistic template functions asymptote to the propagation speed of the field, which for a vacuum is unity.

In the classical~\ac{com} limit, i.e.,~for sufficiently massive detectors~$|\bm{p}| \ll m$,  one re-obtains the results for the traditional~\ac{udw} model. In this regime, the dynamics are independent of the detector's~\ac{com}
\begin{align} \label{eq:classical_1st_template_function}
    \mathcal{T}_{\mathrm{classical}}^{\mathrm{(1st)}}(\bm{p}) &= \frac{E}{\nu} \, ,
\end{align}
and simply proportional to the energy gap. Moreover, in the regime where the detector mass is sufficiently large and dominates all other dimensional quantities, such as $E \ll m$, then all models coincide as seen in Fig.~\ref{fig:template_comparison_vac_E0001}.

In the case of a medium, one can compare the template functions for different propagation speeds~$\nu$ and constant energy gap, as depicted in Fig.~\ref{fig:template_comparison_med}. One observes that the template functions converge for small~$p / m$ (i.e.,~for small momenta or suitably massive detectors) and also finds new transient behavior before converging to the vacuum case. Additionally, while the relativistic first-quantized template function diverges for large~$p / m$, the corresponding semi- and non-relativistic template functions asymptote to the field propagation speed
\begin{align*}
    \lim_{p \to \infty} \mathcal{T}_{\mathrm{semi-rel}}^{\mathrm{(1st)}}(\bm{p}) = \lim_{p \to \infty} \mathcal{T}_{\mathrm{non-rel}}^{\mathrm{(1st)}}(\bm{p}) = \nu \, ,
\end{align*}
which was also observed for the vacuum case in Fig.~(\ref{fig:template_comparison_vac_E0001}).

Due to the presence of the field's propagation speed~$\nu$, one obtains a local maximum in the second-quantized case, most easily seen in Fig.~\ref{fig:template_comparison_med_nu09}, which corresponds to the asymptotic behavior observed for the semi- and non-relativistic~\ac{com} limiting cases. Moreover, this local maximum is still present even for field propagation speeds very close to the speed of light, with large~$\nu$ translating the peaks to larger~$p / m$.


\section{Comparison of spontaneous emission rates}\label{sec:results}
\begin{figure*}
    \centering
    \subfloat[][$L / \lambda_c = 0.1$]{
    \includegraphics[width=0.375\textwidth]{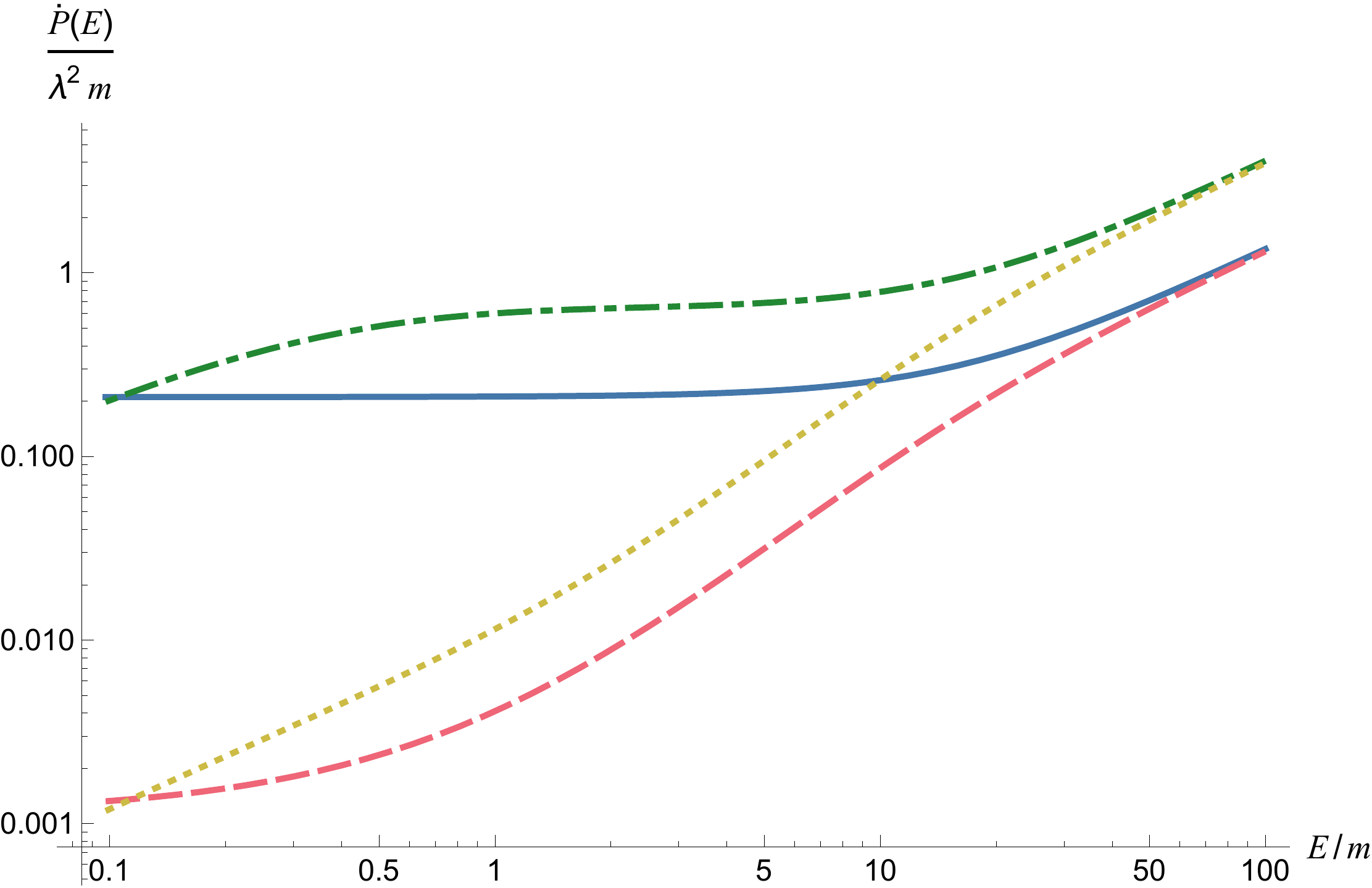}
    \label{fig:transrate_comparison_med_L01}}
    \qquad
    \subfloat[][$L / \lambda_c = 10$]{
    \includegraphics[width=0.375\textwidth]{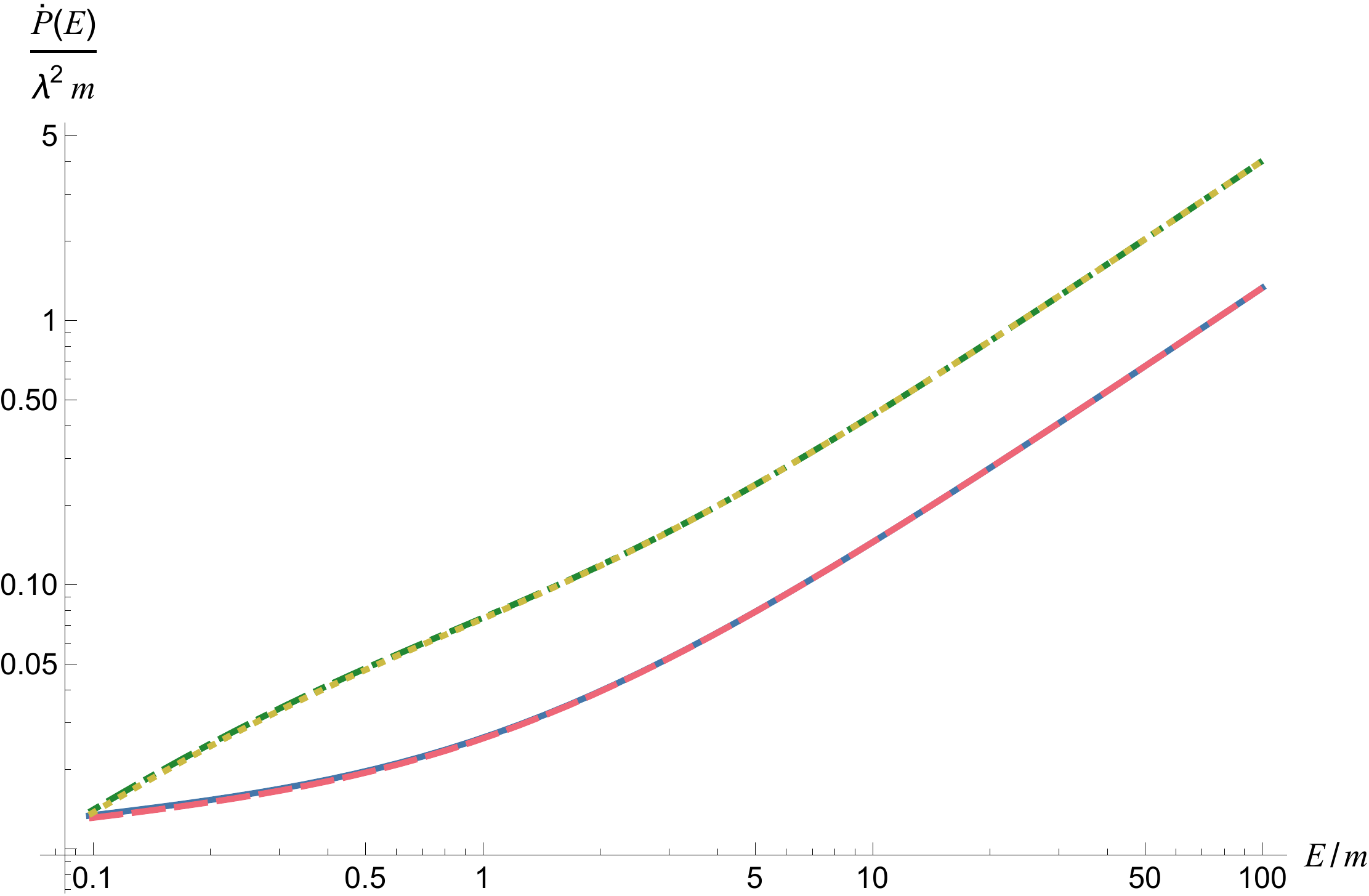}
    \label{fig:transrate_comparison_med_L10}}
    \qquad
    \subfloat{\raisebox{12.5mm}%
    {\includegraphics[width=0.15\textwidth]{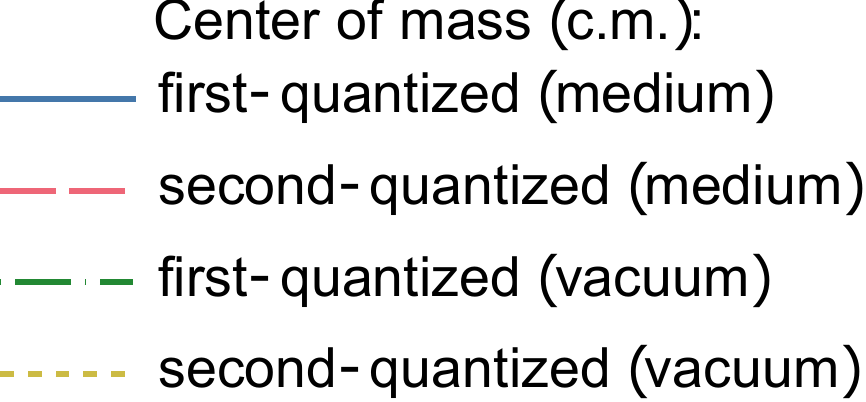}}}
    \caption{Spontaneous emission rates between the relativistic first- and second-quantized localizations for a medium~($\nu = 0.1$) and vacuum~($\nu = c = 1$). Results are given with respect to the Compton scale of the detector, where~$m$ is the detector's rest mass, $\lambda_c \equiv m^{-1}$ the Compton wavelength, and~$L$ the spread of the Gaussian wavefunction in position space. For propagation in a medium, the distributions of the transition rates retain the same form as that in a vacuum, with the two localizations converging for large energy gap~$E / m$, albeit at a lower rate. For (a) Gaussian widths below the Compton wavelength, the first- and second-quantized localizations can be distinguished, although they would not be observable in practice given the inability to empirically access this regime; for (b) Gaussian widths above the Compton wavelength, the two localizations coincide.}
    \label{fig:transrate_comparison_med}
\end{figure*}
\begin{figure*}
    \centering
    \subfloat[][$L / \lambda_c = 0.1$]{
    \includegraphics[width=0.375\textwidth]{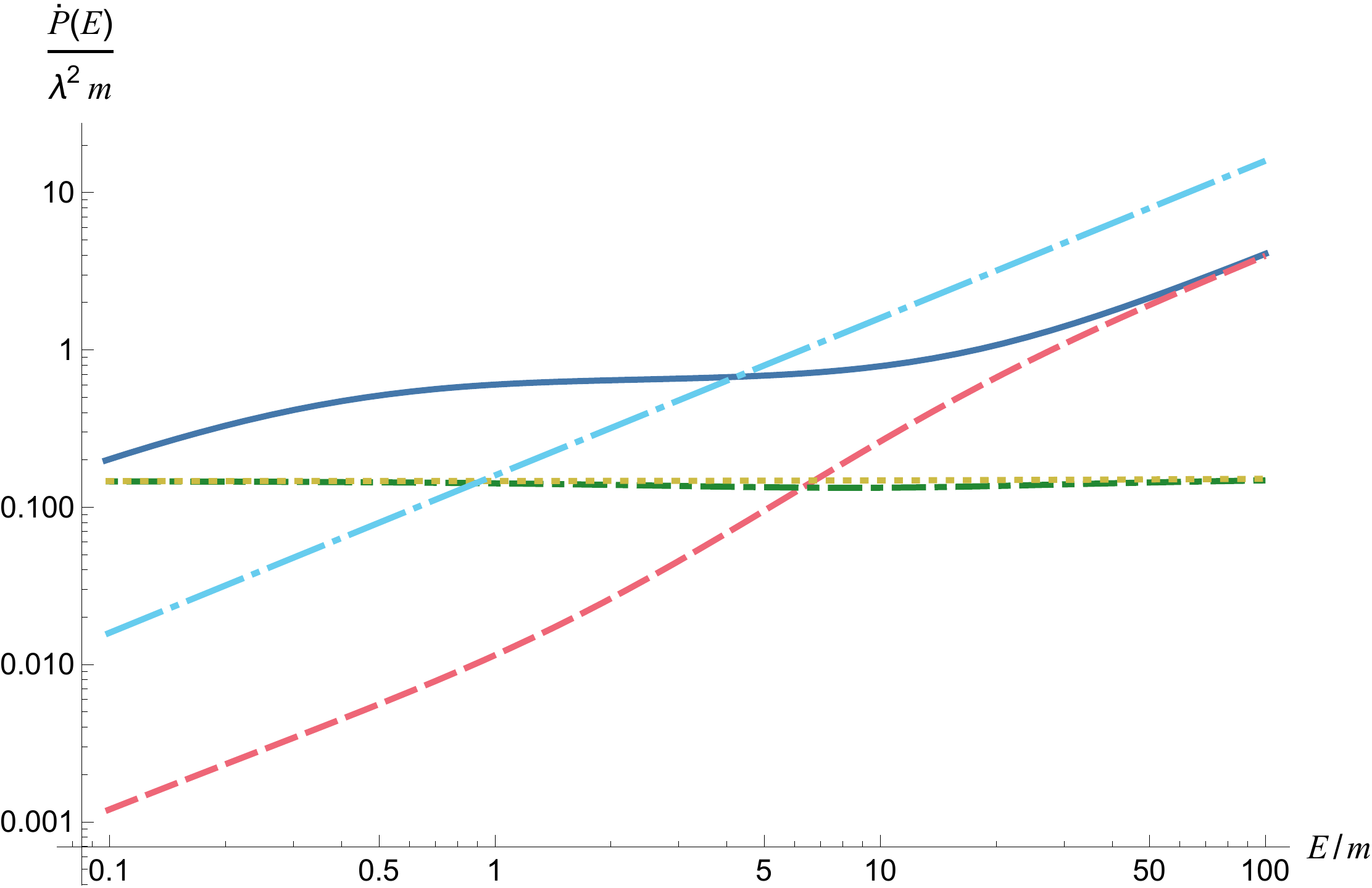}
    \label{fig:transrate_comparison_vac_L01}}
    \qquad
    \subfloat[][$L / \lambda_c = 10$]{
    \includegraphics[width=0.375\textwidth]{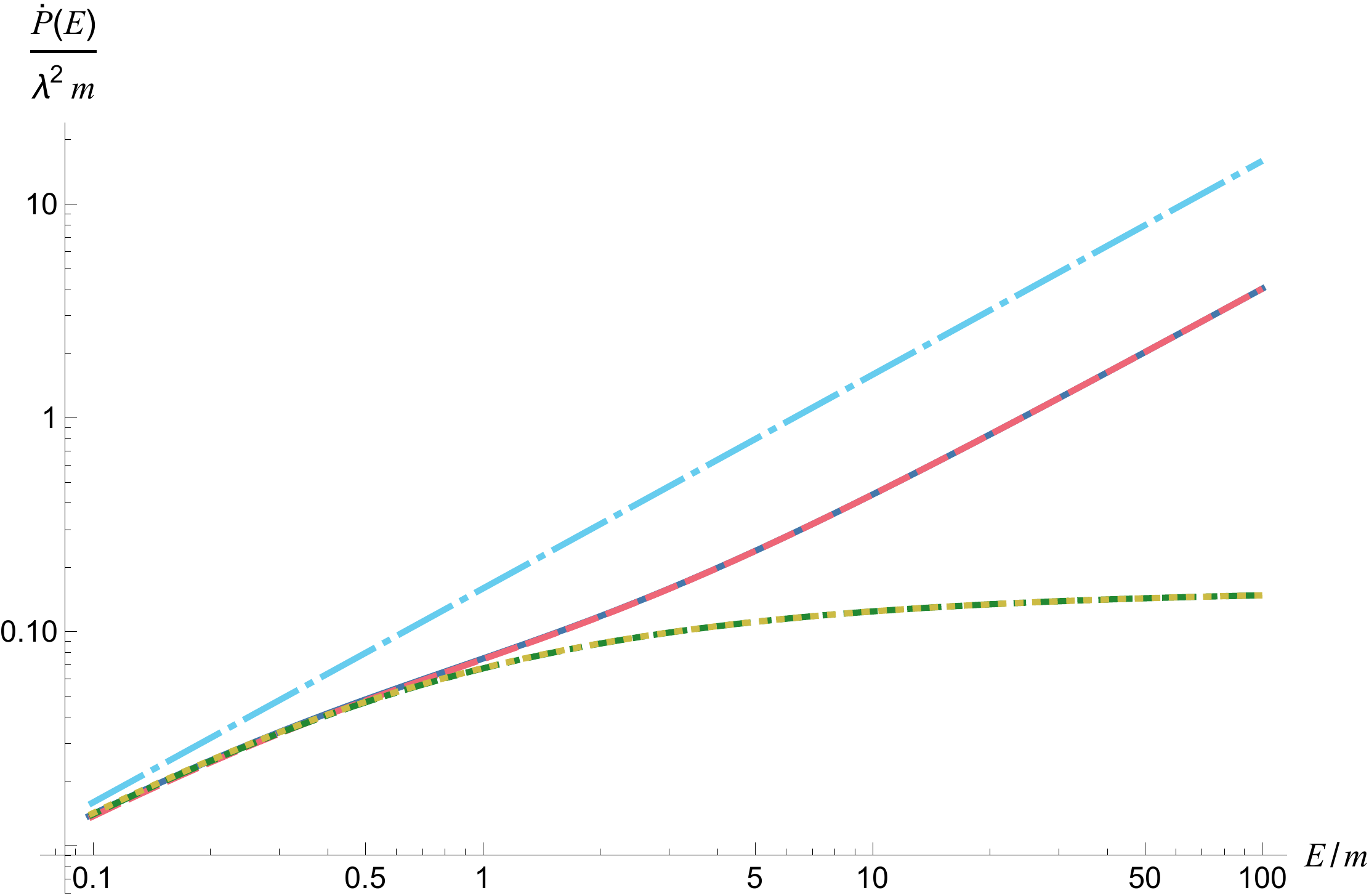}
    \label{fig:transrate_comparison_vac_L10}}
    \qquad
    \subfloat{\raisebox{12.5mm}%
    {\includegraphics[width=0.125\textwidth]{figures/template_comparison_legend.pdf}}}
    \caption{Spontaneous emission rates between the first- and second-quantized localizations in a vacuum, alongside their semi- and non-relativistic limits. Results are given with respect to the Compton scale of the detector, where~$m$ is the detector's rest mass, $\lambda_c \equiv m^{-1}$ the Compton wavelength, and~$L$ the spread of the Gaussian wavefunction in position space. We compare the effect of different Gaussian widths on the transition rate: For small Gaussian widths (a) the relativistic transition rates converge for large energy gap, but do not in the case of a classical~\ac{com}, while the semi- and non-relativistic cases are approximately constant compared to the other cases. For large Gaussian widths (b) the spontaneous emission rates for all cases converge for small energy gaps~$E / m$, and remarkably the relativistic first- and second-quantized localizations coincide for all~$E / m$.}
    \label{fig:transrate_comparison_vac}
\end{figure*}
As defined in the previous section, the transition rate~\eqref{eq:transrate_final} is a functional of the detector's initial wavefunction, and is given by
\begin{align*}
    \dot{P}[\psi_i] \equiv \frac{\lambda^2}{2 \pi} \int d^3p \, |\psi_i(\bm{p})|^2 \, \mathcal{T}(\bm{p}) \, ,
\end{align*}
where the dynamics are entirely described by a respective template function. In the subsequent analysis of the spontaneous emission rate, it is simplest to consider a detector initially in a Gaussian state with width~$L$, which in momentum space is described by
\begin{align} \label{eq:gaussian_wavefunction}
    \psi_i(\bm{p}) = \left( \frac{L^2}{2 \pi} \right)^{3/4} e^{-\frac{L^2}{4} |\bm{p} - \bm{p}_D|^2} \, ,
\end{align}
where the detector is centered about $\bm{p} = \bm{p}_D$.

Analytic results for the relativistic first- and second-quantized cases can be obtained in the case of a vacuum with mean momentum~$\bm{p}_D = \bm{0}$. Substituting the Gaussian wavefunction into the transition rate~(\ref{eq:transrate_final}) alongside the relativistic template functions~(\ref{eq:rel_template_function_compact}), the respective integrals can be evaluated by Euler substitution. For the relativistic first-quantized case, this gives
\begin{subequations} \label{eq:transrates}
\begin{align}
    \dot{P}_{\mathrm{rel}}^{\mathrm{(1st)}}[\psi_i] = \frac{\lambda^2 L \left( M_e^4 - M_g^4 \right)}{4 \pi^{3/2} \sqrt{2} M_e^2} e^{\frac{L^2 M_e^2}{4}} K_1\left( \frac{L^2 M_e^2}{4} \right) \, ,
\end{align}
where~$K_\nu(z)$ is the modified Bessel function of the second kind. In the second-quantized case, one obtains
\begin{align}
    \dot{P}_{\mathrm{rel}}^{\mathrm{(2nd)}}[\psi_i] = \frac{\lambda^2 L \left( M_e^4 - M_g^4 \right)}{4 \pi \sqrt{2} M_e^2} \, U\left( \frac{1}{2},0,\frac{L^2 M_e^2}{2} \right) \, ,
\end{align}
\end{subequations}
where $U(a, b, z)$ is the confluent hypergeometric function of the second kind, and we have expressed the above result in terms of the first-quantized coupling~$\lambda \equiv \lambda^{\mathrm{(1st)}}$ by employing Eq.~(\ref{eq:coupling_relation}).

While analytic results can be easily obtained in the vacuum case, for the semi- and non-relativistic regimes and generally in a medium, the spontaneous emission rate is evaluated numerically. In Fig.~\ref{fig:transrate_comparison_med}, we plot the spontaneous emission rates for the relativistic first- and second-quantized models for both a vacuum and medium. One observes that for the system in a medium, the transition rate is re-scaled when compared to the vacuum case.

Comparing the transitions rates for the relativistic, classical and semi- and non-relativistic~\ac{com} in a vacuum, we analyze the effects of the Gaussian width~$L$ on the results, given in Fig.~\ref{fig:transrate_comparison_vac}. The relativistic localizations converge for large energy gap~$E / m$, but fall short of the value predicted in the case of a classical~\ac{com}. More precisely, in the limit of small rest mass~$m$, one finds that
\begin{align*}
    \lim_{m \to 0} \dot{P}^{\mathrm{(1st)}}_{\mathrm{rel}}[\psi_i] = \lim_{m \to 0} \dot{P}^{\mathrm{(2nd)}}_{\mathrm{rel}}[\psi_i] = \frac{\lambda^2}{2 \pi} \frac{E \nu}{(\nu + 1)^2} \, ,
\end{align*}
while for a classical~\ac{com}, whose template function~(\ref{eq:classical_1st_template_function}) is independent of momentum, the transition rate trivially evaluates to
\begin{align*}
    \dot{P}_{\mathrm{classical}}[\psi_i] = \frac{\lambda^2}{2\pi} \frac{E}{\nu} \, ,
\end{align*}
from which it is clear that the classical and relativistic~\ac{com} disagree in the regime of small mass detectors, which is unsurprising given the mass defect between the ground and excited energy levels. All models coincide for sufficiently massive detectors, i.e.,~when $L / \lambda_c \gg 1$ and $E / m \ll 1$, where the free dynamics of the~\ac{com} become trivial; this regime can be most easily seen in Fig.~\ref{fig:transrate_comparison_vac_L10}, where the models coincide for small energy gap~$E / m$.

In the semi- and non-relativistic regimes, both of the transition rates converge to zero in the limit of small rest mass, that is
\begin{align*}
    \lim_{m \to 0} \dot{P}_{\mathrm{semi-rel}}[\psi_i] = \lim_{m \to 0} \dot{P}_{\mathrm{non-rel}}[\psi_i] = 0 \, .
\end{align*}
As seen in Fig.~\ref{fig:transrate_comparison_vac_L01}, the semi- and non-relativistic results are particularly discrepant in comparison to the relativistic and classical~\ac{com} cases, with significant disagreement for small Gaussian widths, which is most likely due to the use of the Galilean group for the detector~\cite{Note5}.

Notably, in Fig.~\ref{fig:transrate_comparison_vac_L10} for the regime where the Gaussian width is large~$L / \lambda_c \gg 1$, one sees that both localizations for the relativistic~\ac{com} are in close agreement, and perfectly coincide in the infinite width limit~$L \, M_e \to \infty$.
\begin{figure*}
    \centering
    \subfloat[][first-quantized]{
    \includegraphics[width=0.4\textwidth]{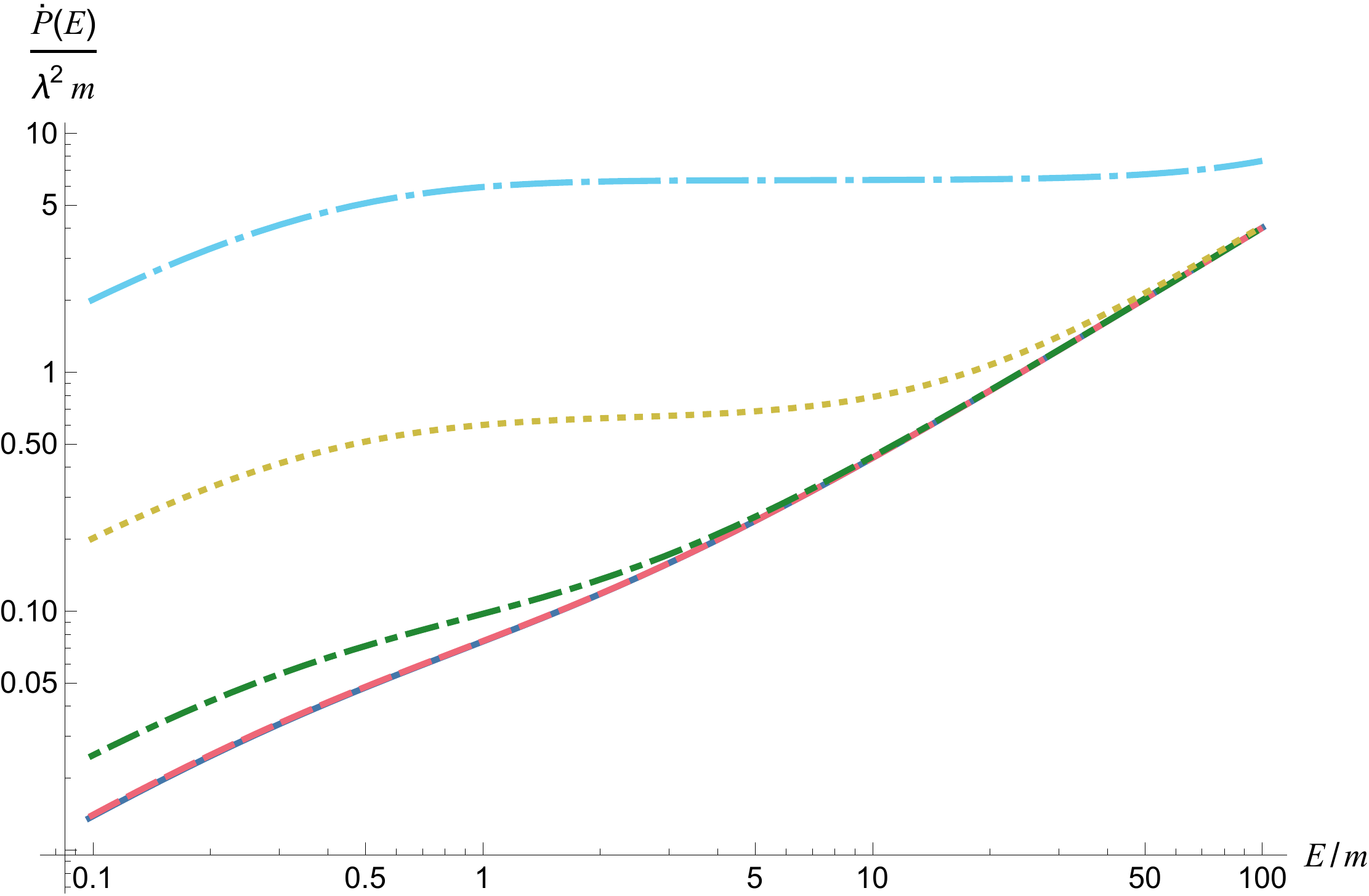}
    \label{fig:transrate_firstq_vac}}
    \qquad
    \subfloat[][second-quantized]{
    \includegraphics[width=0.4\textwidth]{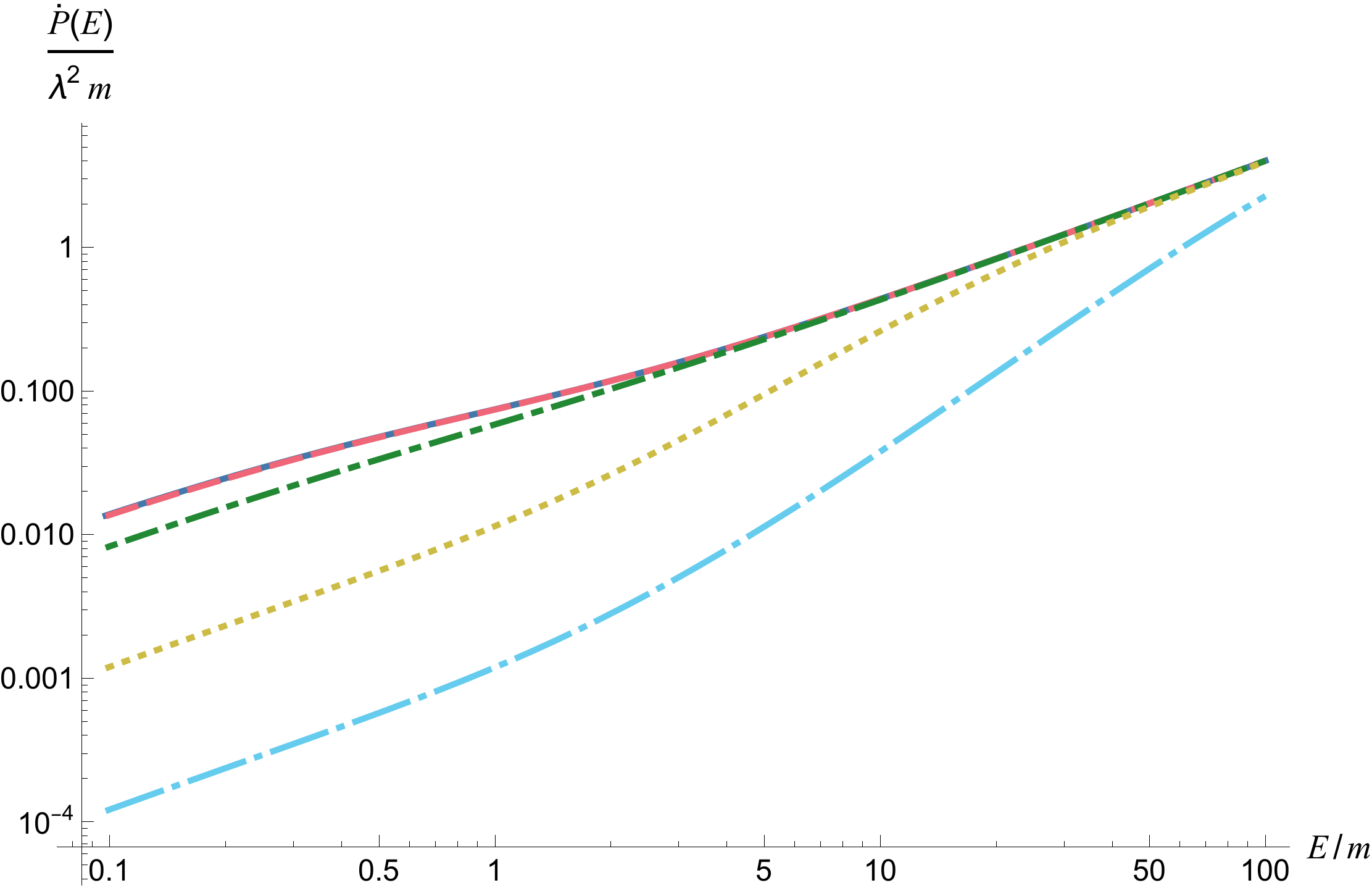}
    \label{fig:transrate_secondq_vac}}
    \qquad
    \subfloat{\raisebox{12.5mm}%
    {\includegraphics[width=0.06\textwidth]{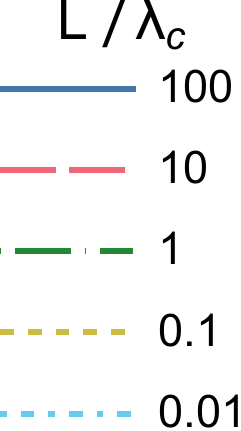}}}
    \caption{Spontaneous emission rates for relativistic (a) first- and (b) second-quantized localizations in a vacuum. Results are given with respect to the Compton scale of the detector, where~$m$ is the detector's rest mass, $\lambda_c \equiv m^{-1}$ the Compton wavelength, and~$L$ the spread of the Gaussian wavefunction in position space. In (a) the first-quantized model, smaller spreads amplify the transition rate with respect to the energy gap of the detector, while the inverse occurs in (b) the second-quantized model, where the rate of spontaneous emission is suppressed for smaller Gaussian widths.}
    \label{fig:transrate_gaussian_spread}
\end{figure*}
Taking this limit in the case of a vacuum, we find
\begin{widetext}
\begin{align*}
    \lim_{L M_e \to \infty} \dot{P}^{\mathrm{(1st)}}_{\mathrm{rel}}[\psi_i] = \frac{\lambda^2 \left( M_e^4 - M_g^4 \right)}{4 \pi M_e^3}\left[ 1 + \frac{3}{2} \frac{1}{(L M_e)^2} + \mathcal{O}\left( \frac{1}{(L M_e)^4} \right) \right] \, ,
\end{align*}
\begin{align*}
    \lim_{L M_e \to \infty} \dot{P}^{\mathrm{(2nd)}}_{\mathrm{rel}}[\psi_i] = \frac{\lambda^2 \left( M_e^4 - M_g^4 \right)}{4 \pi M_e^3}\left[ 1 - \frac{3}{2} \frac{1}{(L M_e)^2} + \mathcal{O}\left( \frac{1}{(L M_e)^4} \right) \right] \, .
\end{align*}
\end{widetext}
To better understand this convergence between the first- and second-quantized localizations, one finds that the two localizations scale inversely with respect to the spread of the Gaussian wavefunction, which can be seen in Fig.~\ref{fig:transrate_gaussian_spread}. For small Gaussian widths, the transition rate for the first-quantized~\ac{com} is amplified, while for the second-quantized~\ac{com} it is suppressed, which ultimately leads to the observed difference in sign between the second-order terms.

Quantitatively, for a hydrogen atom with~$L$ on the order of the Bohr radius, the fractional difference between the two rates is of the order~$\sim 10^{-10}$, which would be very difficult to distinguish empirically; we briefly comment on potential experimental tests in the following section.



\section{Discussion}\label{sec:discussion}

Given the results of the previous sections, a pertinent and obvious question to ask at this stage would be which of the two localizations, the first- or second-quantized, is the ``true'' or ``correct'' one. Due to their different predictions, namely the different emission rates, this question can and should ultimately be resolved by experiment.

Note that the analytical results for the emission rates presented here are for a detector with vanishing mean momentum, which is expected to minimize the difference between these two models. The marked difference in the dependence of the first- and second-quantized template functions on momentum, e.g.,~Eqs.~\eqref{eq:first_quantized_template_function_vacuum} and~\eqref{eq:second_quantized_template_function_vacuum_corrected}, suggests that the predictions of the two models would diverge further for initial states with non-vanishing momentum. Choosing an optimal initial state and exploring other processes, such as absorption, to empirically distinguish between these two models is a worthwhile topic for further research.

Throughout this analysis, we have considered a relativistic first-quantized regime, which some will object to on the grounds that~\ac{rqm} can be consistently formalized only in terms of a relativistic~\ac{qft}, i.e.,~the second-quantized model outlined in Sec.~\ref{sec:second_quantized}. Such an argument is often made on the basis of relativistic covariance, namely that both space and time must be treated equivalently under the Lorentz group. While this is an important point, it must be stressed that the issue of localization exists independently of covariance. Both a clear physical explanation of the discrepancy between the two localizations, and a theoretical argument for which of the predictions would correspond to empirical observation, are missing. Filling this gap is necessary for a complete understanding of~\ac{rqm}.

Regarding which of the two localizations provides the ``correct'' description, there are a number of different positions one can take, of which we discuss the following:
\begin{enumerate}[label=(\roman*)]
    \item The ``second-quantized'' localization provides a better representation of the physics, since a complete account of a relativistic theory must ultimately be formalized in terms of quantum fields.
    \item The ``first-quantized'' localization is an idealization, which provides an effective model of the dynamics within an appropriate regime.
    \item The ``first-quantized'' localization provides a better representation, even beyond an effective description. Such a localization would presumably play a fundamental role in relativistic quantum mechanics, particularly in its connection to an appropriately defined covariant position operator.
\end{enumerate}
Considering these different positions, one might intuitively take~(i) to be obviously correct. However, such a view would seriously challenge a particle detector description, such as the standard~\ac{udw} model. The introduction of a ``particle detector'' was motivated precisely to avoid these issues regarding both the definitions of a ``particle'' and localization! By introduction of a ``particle detector'', one couples a system of relativistic quantum fields to a detector, which necessarily assumes a localized system in contradiction to~(i).

Perhaps, to avoid abandoning the detector concept, one could adopt the viewpoint given by~(ii). Thus, a localized description only provides an effective model, and one would interpret the detector as an idealized system,~e.g.,~a relativistic composite system such as an atom, where the degrees of freedom of the model are not fundamental, implying a coarse-grained~\ac{com}. By characterizing a localized system through the introduction of a position operator, one obtains an effective description of the underlying fundamental fields, at least according to~(ii).

Arguably, however,~(ii) merely defers the problem. Rather than treat the ``detector'' as an effective model of an atom, one could in principle equally apply the model to an elementary system, wherein the detector is interpreted as an electron or neutrino. By this view, one is seemingly led to~(iii), which itself has no easy interpretation, and one must again confront the question of localization in relativistic quantum mechanics.


\subsection{Position operator and localization}\label{sec:localization}

The problem of localization in~\ac{rqm} has a long history, with its initial study originating in the seminal paper by~\citet{pryce_mass-centre_1948}, covering the different possible definitions of a relativistic~\ac{com}. As opposed to the Newtonian case under the Galilean group, one finds many different competing definitions of a~\ac{com} in relativity, whether it be classical or quantum. Shortly after Pryce's analysis,~\citet{newton_localized_1949} investigated possible definitions for a relativistic position operator by imposing a number of invariance conditions, from which they derived their Newton-Wigner position operator, which is equivalent to the standard position operator of~\ac{nrqm}.

Subsequent analyses (see, e.g.,~\cite{kalnay_localization_1971} and references within) have further investigated the problem of localization, namely whether one can introduce a covariant position operator and whether localized states are consistent with~\ac{rqm} and requirements of causality. The~\ac{udw} model provides an especially simple and convenient theoretical framework to study this question, particularly the consequences of what we refer to as the first- and second-quantized localizations. Moreover, in the massive spin-0 case, the different possible definitions for a relativistic~\ac{com} are comparatively simpler and in this instance equivalent~\cite{jordan_lorentz-covariant_1963, fleming_covariant_1965}, which is particularly convenient for studying and interpreting the first- and second-quantized localizations.

In the first-quantized model, we required that the respective position states be identical to that of~\ac{nrqm}, which is equivalent to imposing the Newton-Wigner localization and defining a corresponding position operator that obeys~$\hat{\bm{x}}_{\scriptscriptstyle\mathrm{NW}} \ket{\bm{x}}_D = \bm{x} \ket{\bm{x}}_D$~\cite{newton_localized_1949}. As discussed previously in Sec.~\ref{sec:second_quantized}, by imposing this localization scheme one obtains orthogonal position states, unlike that obtained for the second-quantized model. The difficulty of these second-quantized ``position states''~(\ref{eq:second_quantized_position_state}) is that they are non-orthogonal, which can be most easily seen by taking the inner product
\begin{align*}
    \braket{\bm{x}_j | \bm{y}_k}_D^{(\mathrm{2nd})} = \int \frac{d^3p}{2 (2\pi)^3 \sqrt{E_j(\bm{p}) E_k(\bm{p})}} e^{i \bm{p} \cdot (\bm{x} - \bm{y})} \delta_{jk} \, ,
\end{align*}
with indices denoting the internal degrees of freedom. Working in spherical coordinates, and suppressing indices, the above integral can be evaluated to
\begin{align*}
    \braket{\bm{x} | \bm{y}}_D^{(\mathrm{2nd})} &= \frac{M}{(2 \pi)^2 |\bm{x} - \bm{y}|} K_1 \left( M |\bm{x} - \bm{y}| \right) \, ,
\end{align*}
which can alternatively be rewritten in terms of the Compton wavelength~$\lambda_c = 1 / M$ with respect to the detector mass~$M$
\begin{align*}
    \braket{\bm{x} | \bm{y}}_D^{(\mathrm{2nd})} &= \frac{1}{(2 \pi)^2 \lambda_c |\bm{x} - \bm{y}|} K_1 \left( \frac{1}{\lambda_c} |\bm{x} - \bm{y}| \right) \, .
\end{align*}
Following~\cite{padmanabhan_obtaining_2018}, the dependence on the Compton scale can be more easily seen by Taylor expanding the integration measure~$E(\bm{p}) = \sqrt{\bm{p}^2 + M^2}$ about~$|\bm{p}| = 0$
\begin{align*}
    \braket{\bm{x} | \bm{y}}_D^{(\mathrm{2nd})} &= \frac{1}{2 M} \int \frac{d^3p}{(2\pi)^3} \left( 1 - \frac{\bm{p}^2}{2 M^2} + \mathcal{O}(\bm{p}^4) \right) e^{i \bm{p} \cdot (\bm{x} - \bm{y})} \\
    &= \frac{1}{2 M} \left( 1 + \frac{1}{2} \lambda_c^2 \bm{\nabla}^2 + \mathcal{O}(\lambda_c^4) \right) \delta^{(3)}(\bm{x} - \bm{y}) \, ,
\end{align*}
where one finds that the position states are orthogonal up to second order in the Compton wavelength. The first- and second-quantized localizations are equivalent provided that the Compton scale is sufficiently small with respect to spread of the wavefunction in position space, which was observed in the~\ac{udw} model for large Gaussian widths in Fig.~\ref{fig:transrate_comparison_vac_L10}.

\subsection{\texorpdfstring{\acf{fw}}{Foldy-Wouthuysen (FW)} transformation}\label{sec:foldy_wouthuysen}

The results of~\citet{newton_localized_1949} were subsequently extended by~\citet{foldy_dirac_1950}, who found the representation of the Newton-Wigner position operator for the spin-1/2 case when the positive and negative energy states are decoupled. Moreover, Foldy and Wouthuysen found their eponymous unitary transformation relating the two representations, where the decoupled representation gives a Newton-Wigner position operator identical to the non-relativistic position operator~\cite{costella_foldywouthuysen_1995}.

In the relativistic second-quantized~\ac{udw} model, the non-orthogonal position states are transformed into the first-quantized position states by the action of an~\ac{fw} transformation, which maps between the respective first- and second-quantized localizations, and decouples the internal states of the detector. Moreover, by transforming to the first-quantized representation, one loses manifest Lorentz covariance but obtains a position operator with corresponding position eigenstates.

While in the literature it is common to take the second-quantized localization as more fundamental, and thus as more correct in any relativistic context, it may be premature here. Note that we are concerned with an effective description of a composite system, such as an atom, whose many degrees of freedom are reduced to just the~\ac{com} and two internal states; taking such an effective model as fundamental by treating it within~\ac{qft} appears inconsistent. Indeed, in scenarios like ours, the predictions from the first-quantized model may have a greater credence than those obtained by modeling the complexities of an atom as a two-component scalar field.



Further theoretical work combined with experiments is necessary to distinguish between the first- and second-quantized localizations and their corresponding representations. In particular, while the~\ac{fw} transformation is commonly interpreted as a purely mathematical transformation which decouples the internal states, it may be possible to give it a physical interpretation, and understand its role both in mapping between the two different localizations and in the loss of covariance that results from the first-quantized representation.

\subsection{Time and covariance}\label{sec:time_covariance}

In the traditional~\ac{udw} model, one conventionally parameterizes the classical trajectory of the detector by its proper time; however, in the case of a relativistic detector with a quantized~\ac{com}, the treatment of time is conceptually more difficult, due to the various ``problems of time'' one encounters in~\ac{rqm}. One particular strategy to formalizing a first-quantized~\ac{rqm} in this way proceeds by introducing an additional invariant evolution parameter, such as a proper-time, which parameterizes the spacetime coordinates. These proper-time dynamical formulations were initially developed in the early twentieth century by a number of eminent physicists, and have since been developed considerably (for a review of the different approaches, see~\cite{fanchi_review_1993} and references within); one of the most recent realizations of a proper-time dynamical formulation being the~\ac{shp} theory~\cite{stuckelberg_signification_1941,*stuckelberg_remarque_1941,*stuckelberg_mecanique_1942, horwitz_relativistic_1973, horwitz_relativistic_2015}. While it has been straightforward to consider the inertial dynamics of an~\ac{udw} detector in the present formulation developed here, a proper-time or path-integral formulation would be beneficial for the study of more general trajectories, namely in the case of non-inertial detectors and curved spacetimes.

Closely related to the aforementioned problems of localization and time is the difficulty in defining a covariant position operator and state with respect to the Lorentz transformations. In conventional~\ac{nrqm}, one can identify a wavefunction in the position basis parametrized by time, and obtain a covariant wavefunction under the Galilean group. However, the relativistic extension is highly non-trivial due to the mixing of space and time under Lorentz transformations, and problematizes a relativistic first-quantized approach. This is avoided in a relativistic~\ac{qft}, where instead one defines a field operator as a function of spacetime, which is manifestly covariant and bypasses this particular problem of time in a first-quantized formalism. 

\subsection{Mass-energy operator}\label{sec:mass_energy}

In traditional models, mass-energy (mass) is treated as a $c$-number and is formally described as the central charge of the Poincar\'{e} (Galilean) group. In many relativistic extensions to quantum mechanics, one considers a central extension of the underlying group structure and treats mass as a dynamical variable~\cite{horwitz_relativistic_2015, greenberger_theory_1970,*greenberger_theory_1970-1}, although we do not consider such a treatment here. In our relativistic first-quantized model, we promote mass-energy to an operator, such that the dispersion relation for the second-quantized detector with two mass components~(\ref{eq:detector_second_q}) has in flat spacetime the same dispersion relation given by Eq.~(\ref{eq:rqm_free}) for a mass operator defined by Eq.~(\ref{eq:mass_energy_operator_two_level}).

The motivations for introducing a mass-energy operator are numerous, and it has been introduced multiple times independently in the literature. A mass-energy operator naturally follows from proper-time dynamical treatments of~\ac{rqm}~\cite{johnson_position_1969, fanchi_review_1993, zych_quantum_2011, pikovski_universal_2015, smith_quantum_2020}; or relatedly, within a dynamical mass approach, it can be obtained through canonical quantization by introducing a conjugate variable with respect to the mass~\cite{giulini_galilei_1996, annigoni_mass_2013}. More recently, studies of the Einstein equivalence principle in the quantum regime have also motivated the introduction of quantized mass-energy~\cite{zych_quantum_2018, tobar_mass-energy_2022}; and finally, a compelling empirical motivation for introducing a mass operator is the discovery of neutrino oscillations, which has found an explanation in terms of superpositions of mass eigenstates~\cite{fantini_introduction_2017}.


\section{Conclusion}\label{sec:conc}

In this paper, we proposed and studied an extension of the~\ac{udw} detector model which incorporates a first-quantized relativistic~\ac{com}. By studying the process of spontaneous emission, we have found that the transition rates are different between our first-quantized description and the second-quantized model, and likewise for the cases of a classical, semi- and non-relativistic~\ac{com}. Furthermore, we have studied the detector-field system in both a vacuum and medium, obtaining analytic results in the former case.

The different predictions between the relativistic models are ultimately due to the different localization schemes implied by  the first- and second-quantized approaches. Significantly, for these two cases we find that the dynamics encapsulated by the ``template functions'' are notably distinct, which suggests that the two localizations can be empirically studied. In the interest of future experiments, these results can and should be considered for other physical processes such as absorption and stimulated emission; likewise, one should also search for optimal detector states in order to find more easily testable parameter regimes.

On the theoretical side, we indicate that the two localizations are related by an~\ac{fw} transformation, the physical role of which must still be clarified given the importance of its mapping between the two schemes. Extending the model to include the spin of the detector and field would also be beneficial for future investigations, since the inclusion of angular momentum in the model may lead to additional empirically distinguishable effects resulting from the two localizations.

For the models we have developed in this paper, we have only considered inertial detectors in Minkowski spacetime. It is of obvious interest to extend this study to non-inertial trajectories and curved spacetimes, of which the former can be modeled by introducing an electromagnetic field to accelerate (e.g.,~a charged) detector. One could then study the impact of the detector's localization on the Unruh effect and further explore the~\ac{fw} transformation in the context of curved spacetimes.

Lastly, while one can provide a relativistically covariant description for the detector following a second-quantized approach, it is worthwhile exploring whether a first-quantized model can also be formalized covariantly. Such a project could be developed via a proper-time dynamical formulation, which can be developed in several ways, such as in terms of a path-integral formulation or alternatively in terms of a relativistic dynamical approach, e.g.,~using the~\ac{shp} theory. Such an investigation can help to clarify the relation between the Lorentz and~\ac{fw} transformations, and may have interesting implications for the study of quantum reference frames.

\begin{acknowledgments}
We thank Carolyn E. Wood, Fabio Costa, Joshua Foo, Laura J. Henderson, Valentina Baccetti and Nicolas C. Menicucci for helpful discussions and advice. We additionally thank Timothy C. Ralph and Robert B. Mann for pressing us on the proper interpretation of our results, pertaining to the differences between the first- and second-quantized localizations.

This work was supported by the ARC Future Fellowship Grant No.~FT210100675 and through the ARC Centre of Excellence for Engineered Quantum Systems (EQUS) Grant No.~CE170100009.

The authors acknowledge the Turrbal and Jagera peoples, who are the traditional owners of the land on which the University of Queensland is situated. We pay our respects to their Elders, past and present.
\end{acknowledgments}


\appendix*

\section{Scalar field, conventions, and derivation}\label{app:scalar_field}

In the interest of providing a self-contained analysis, we derive the field operator for a scalar field following similar treatments given by~\citet{birrell_quantum_1982} and~\citet{peskin_introduction_2019}, employing the same conventions commonly used in the~\ac{udw} literature. As opposed to the main body of the paper, here we consider a $(d + 1)$-dimensional Minkowski spacetime.

A free real-valued scalar field $\phi$ has Lagrangian density
\begin{align}
    \mathcal{L}[\phi, \partial_\mu \phi] = \frac{1}{2} \left( \partial^\mu \phi \partial_\mu \phi - m^2 \phi^2 \right) \, ,
\end{align}
from which one can derive the Klein-Gordon equation
\begin{align}
    \left( \Box + m^2 \right) \phi(x) = 0 \, ,
\end{align}
where $\Box \equiv \partial^\mu \partial_\mu$ denotes the d'Alembert operator and $m$ is the mass. The field equation is solved by a set of solutions for positive and negative frequencies~$u^{\scriptscriptstyle (\pm)}_{\bm{k}}$ with momentum~$\bm{k}$. Following convention, we choose a planewave basis for the field modes such that
\begin{align}
    u^{\scriptscriptstyle (\pm)}_{\bm{k}}(x) = \frac{1}{f(\bm{k})} e^{\mp i k \cdot x} \, ,
\end{align}
where $k \cdot x = \omega(\bm{k}) t - \bm{k} \cdot \bm{x}$, $\omega(\bm{k}) = \sqrt{\bm{k}^2 + m^2}$ and $f(\bm{k})$ is an arbitrary normalization factor. For two solutions~$\phi_j$ and~$\phi_k$ of the Klein-Gordon equation, one obtains the conserved current
\begin{align}
    J^\mu_{jk} = i \left( \phi_j^* \partial^\mu \phi_k - \phi_k \partial^\mu \phi_j^* \right ) \, ,
\end{align}
and with respect to the time-component, one defines the Klein-Gordon inner product
\begin{align}
    \left( \phi_j, \, \phi_k \right) = i \int d^dx \, \left( \phi_j^* \partial_t \phi_k - \phi_k \partial_t \phi_j^* \right) \, .
\end{align}
To ensure orthonormal~$u_{\bm{k}}$ modes, we employ the symmetric Fourier convention~$(2\pi)^{-d/2}$ and non-covariant integration measure~$(2\omega(\bm{k}))^{-1/2}$, such that
\begin{align}
    u_{\bm{k}}(x) = \frac{1}{(2\pi)^{d/2} \sqrt{2 \omega(\bm{k})}} e^{-i k \cdot x} \, ,
\end{align}
where $u_{\bm{k}} \equiv u^{\scriptscriptstyle (+)}_{\bm{k}}$ and $u^*_{\bm{k}} \equiv u^{\scriptscriptstyle (-)}_{\bm{k}}$. A general solution to the Klein-Gordon equation can be constructed by decomposing the field into its positive and negative frequency components
\begin{align}
    \phi(x) = \phi^{\scriptscriptstyle (+)}(x) + \phi^{\scriptscriptstyle (-)}(x) \, ,
\end{align}
where~$\phi^{\scriptscriptstyle (\pm)}$ can in turn be expressed as a superposition of the corresponding $u^{\scriptscriptstyle (\pm)}_{\bm{k}}$ modes, such that
\begin{align}
    \phi(x) = \int d^dk \left( a^{\scriptscriptstyle (+)}(\bm{k}) u^{\scriptscriptstyle (+)}_{\bm{k}}(x) + a^{\scriptscriptstyle (-)}(\bm{k}) u^{\scriptscriptstyle (-)}_{\bm{k}}(x) \right) \, .
\end{align}
The field $\phi$ can be quantized by introducing a canonically conjugate field
\begin{align}
    \pi(x) = \frac{\partial \mathcal{L}(x)}{\partial \bm{(} \partial_t \phi(x) \bm{)}} = \partial_t \phi(x) \, ,
\end{align}
such that $\phi$ and $\pi$ are promoted to operators which obey the equal-time canonical commutation relations
\begin{subequations}
\begin{align}
    & [\hat{\phi}(\bm{x}, t), \hat{\pi}(\bm{x}', t)] = i \delta^{(d)}(\bm{x} - \bm{x}') \, , \\
    & [\hat{\phi}(\bm{x}, t), \hat{\phi}(\bm{x}', t)] = [\hat{\pi}(\bm{x}, t), \hat{\pi}(\bm{x}', t)] = 0 \, .
\end{align}
\end{subequations}
The field operator $\hat{\phi}$ may now be expanded in terms of the field modes
\begin{align}
    \hat{\phi}(x) = \int d^dk \left( \hat{a}(\bm{k}) u_{\bm{k}}(x) + \hat{a}^\dagger(\bm{k}) u^*_{\bm{k}}(x) \right) \, ,
\end{align}
where~$\hat{a}^\dagger(\bm{k})$ and~$\hat{a}(\bm{k})$ are respectively the creation and annihilation operators, which obey the non-covariant commutation relations
\begin{subequations}
\begin{align}
    & [\hat{a}(\bm{k}), \hat{a}^\dagger(\bm{k}')] = \delta^{(d)}(\bm{k} - \bm{k}') \, , \\
    & [\hat{a}(\bm{k}), \hat{a}(\bm{k}')] = [\hat{a}^\dagger(\bm{k}), \hat{a}^\dagger(\bm{k}')] = 0 \, .
\end{align}
\end{subequations}
The action of an annihilation operator on the vacuum state is defined to be
\begin{align}
    \hat{a}(\bm{k}) \ket{0} = 0 \, , \; \forall \bm{k}
\end{align}
and a creation operator acting on the vacuum gives
\begin{align}
    \hat{a}^\dagger(\bm{k}) \ket{0} = \ket{\bm{k}} \, ,
\end{align}
which defines a single-particle state with momentum $\bm{k}$. Expanding out the field operator in full, one obtains
\begin{widetext}
\begin{equation} \label{eq:app_general_soln}
    \hat{\phi}(\bm{x}, t) = \int \frac{d^dk}{(2\pi)^{d/2} \sqrt{2 \omega(\bm{k})}} \left( \hat{a}(\bm{k}) e^{-i (\omega(\bm{k}) t - \bm{k} \cdot \bm{x})} + \hat{a}^\dagger(\bm{k}) e^{i (\omega(\bm{k}) t - \bm{k} \cdot \bm{x})} \right) \, .
\end{equation}
One can derive the dimensions of $\phi$ most easily by comparing the Lagrangian density and its kinetic term $[\mathcal{L}] = [(\partial_\mu \phi)^2]$, which implies that $[\phi] = \sqrt{[\mathrm{energy}] / [\mathrm{length}]^{d - 2}}$, and restoring $\hbar$ and $c$ gives
\begin{equation}
    \hat{\phi}(\bm{x}, t) = \int \frac{d^{d}k}{(2\pi)^{d/2}} \sqrt{\frac{\hbar c^2}{2 \omega(\bm{k})}} \left( \hat{a}(\bm{k}) e^{-i (\omega(\bm{k}) t - \bm{k} \cdot \bm{x})} + \hat{a}^\dagger(\bm{k}) e^{i (\omega(\bm{k}) t - \bm{k} \cdot \bm{x})} \right) \, .
\end{equation}
For a field in a medium, one replaces the speed of light~$c$ with the medium's propagation speed~$\nu$, as given by Eq.~\eqref{eq:field_operator}.
\end{widetext}

\footnotetext[1]{For localized quantum systems (i.e.,~detectors) coupled locally to a quantum field, entanglement can be transferred between the field and detectors. Consequently, the entanglement present in the state of the quantum field can be treated operationally. Such a scenario, where entanglement is transferred from the field to detectors, is referred to as ``entanglement harvesting'', see, e.g.,~\cite{salton_acceleration-assisted_2015, martin-martinez_sustainable_2013}.}
\footnotetext[2]{A caveat of this result being that, while the correlation functions converge, the norms do not~\cite{colosi_what_2008}.}
\footnotetext[3]{Unlike many instances where this measure appears, the presence here of this integration measure is not merely a choice of convention. It cannot be eliminated by a redefinition without a corresponding rescaling in momentum space, and consequently redefining the commutation relations between the creation and annihilation operators.}
\footnotetext[4]{Note that here a state is a collection of  position amplitudes for all points in space at some fixed time. However, when viewed from a different reference frame, this state is a collection of position amplitudes at different times, i.e.,~not a state in that frame.}
\footnotetext[5]{For small Gaussian widths, contributions from high momenta to the dynamics are increasingly relevant, which leads to the observed differences between the transition rates for the relativistic detectors, and for the non- and semi-relativistic models.}

\bibliography{references}

\end{document}